\documentclass[12pt]{iopart}
\pdfoutput=1
\usepackage{iopams}
\expandafter\let\csname equation*\endcsname\relax
\expandafter\let\csname endequation*\endcsname\relax
\usepackage[square,numbers,sort&compress]{natbib}
\usepackage{amssymb,color}

\usepackage{bm} 
\usepackage{color}
\usepackage{graphicx}
\usepackage{hhline}
\usepackage{physics}
\usepackage{amsthm}
\usepackage[section]{placeins}
\usepackage{layouts}
\usepackage{amssymb}
\usepackage{enumerate}
\usepackage{placeins}
\usepackage{float}
\usepackage{booktabs}
\usepackage{multirow}
\usepackage{dsfont}
\usepackage{yfonts}
\usepackage{hyperref} 

\DeclareSymbolFont{bbold}{U}{bbold}{m}{n}
\DeclareSymbolFontAlphabet{\mathbbold}{bbold}

\usepackage[color=blue!60!white!40!]{todonotes}

 \newcommand{\hypergeometric}{{{}_1}\hspace{-0.06cm}F\!{{}_2}}

\newcommand{\rhonocl}{\tilde{\rho}_{1}}

\newcommand{\ren}{R\'{e}nyi~}

\begin{document}

\title{A Scaling Function for the Particle Entanglement Entropy of Fermions }
\author{Harini Radhakrishnan}
\address{Department of Physics and Astronomy, University of Tennessee, Knoxville, TN 37996, USA}

\address{Institute for Advanced Materials and Manufacturing, University of Tennessee, Knoxville, TN 37996, USA}

\author{Matthias Thamm}
\address{Institut f\"{u}r Theoretische Physik, Universit\"{a}t Leipzig,  Br\"{u}derstrasse 16, 04103 Leipzig, Germany}

\author{Hatem Barghathi}
\address{Department of Physics and Astronomy, University of Tennessee, Knoxville, TN 37996, USA}
\address{Institute for Advanced Materials and Manufacturing, University of Tennessee, Knoxville, TN 37996, USA}

\author{Bernd Rosenow}
\address{Institut f\"{u}r Theoretische Physik, Universit\"{a}t Leipzig,  Br\"{u}derstrasse 16, 04103 Leipzig, Germany}

\author{Adrian Del Maestro}
\address{Department of Physics and Astronomy, University of Tennessee, Knoxville, TN 37996, USA}
\address{Min H. Kao Department of Electrical Engineering and Computer Science, University of Tennessee, Knoxville, TN 37996, USA}
\address{Institute for Advanced Materials and Manufacturing, University of Tennessee, Knoxville, TN 37996, USA}

\date{\today}

\begin{abstract}
    Entanglement entropy under a particle bipartition provides complementary information to mode entanglement as it is sensitive to interactions and particle statistics at leading order and does not depend on any externally imposed length scale. 
    In this paper, we investigate the particle entanglement entropy in a system of $N$ interacting spinless lattice fermions in one spatial dimension by combining bosonization techniques with exact and approximate numerical methods. We introduce a general scaling form for the fermionic particle entanglement entropy captured by a shape function that enters as an extensive interaction induced correction to a known free fermion result. A general asymptotic expansion in the total number of particles demonstrates that its form is robust for different values of the R{\'e}nyi index and highlights how quantum correlations are encoded in the $n$-particle density matrix of a pure many-body quantum state. 
  
\end{abstract}
 
\maketitle
\newpage

\section{Introduction}
Any $n$-body observable of a quantum $N$-particle system can be calculated from its $n$-particle reduced density matrix ($n$-RDM) without the need to construct the complete wavefunction. The spectrum of the $n$-RDM contains explicit information on non-classical correlations, and its von Neumann or \ren entropy can be used to quantify the quantum entanglement between partitions of $n$ and $N-n$ indistinguishable particles \cite{Zanardi:2002xc,Shi:2003ro,Zozulya:2008bg,Haque:2009zi,Bulgac2022Nuclear}. In one spatial dimension (1D) this \emph{particle} entanglement entropy $S(n)$ can be sensitive to interactions, phase transitions \cite{wu2004quantum, wei2018linking}, and particle statistics at leading order. It has been used to probe quantum Hall states \cite{Haque:2007mu,Iblisdir:2007,Zozulya:2007fe,Pu:2022pe,Hudomal:2019,Liu:2010pe} as well as systems of non-interacting \cite{Simon:2002}, and interacting bosons \cite{Herdman:2014jy,Herdman:2014vd,Herdman:2015xa,Katsura:2007ty,Mohd2022}, fermions \cite{Barghathi:2017tg,Rammelmueller:2017al,Iemini:2015ze,Ferreira:2022qq}, and anyons \cite{Santachiara:2007it}.  

Thus, the particle entanglement can provide complementary information to the well studied and more conventional \emph{mode} entanglement entropy most commonly computed from a reduced density matrix of spatial modes. For example, in 1D, spatial mode entanglement entropy is known to be universal at leading order in the subsystem size $\ell$ for gapless systems describable by a conformal field theory with central charge $c$ \cite{Calabrese:2004ll}, \emph{i.e.}\@ $S(\ell) \sim ({c}/{3}) \ln \ell + \dots$ for the von Neumann entropy.  In contrast, it was empirically proposed by Haque et al.\@ \cite{Haque:2007mu} that the particle entanglement entropy behaves as $S(n) \sim n \ln N$ for $n \ll N$ and $N \gg 1$ for fermions. This was subsequently confirmed for the 2nd R\'enyi entropy within the Luttinger liquid framework for $n=1$ by a subset of the authors of this work \cite{Barghathi:2017tg}.  For bosons in 1D described by Luttinger liquid theory, it was found that for $n=1$, the leading term may become non-universal \cite{Herdman:2015xa}. Existing results are all consistent with Coleman's theorem \cite{coleman1963structure} stating that the $1$-particle von Neumann entropy is minimal for a Slater determinant: $S(n=1) \ge \ln N$. For $n>1$ exact results are scarce, and it was conjectured in 2016 \cite{Carlen:2016} that for $n=2$, $S(n=2) \ge \ln \binom{N}{2}$ which was adapted in Ref.~\cite{lemm2017entropy} to general $n$: $S(n) \ge \ln N$, demonstrating that all possible particle bipartitions of a $N$ fermion system are entangled, a rather striking result! In spite of this recent progress, a more complete understanding of the particle entanglement entropy as a function of $n$, $N$, or R\'enyi index remains elusive. 

Here, we address this question by considering a system of interacting spinless lattice fermions in 1D through a combination of bosonization techniques for $n=1$ and large scale exact diagonalization (ED) and density matrix renormalization group (DMRG) for general $n$.  By exploiting all symmetries of the $n$-RDM, we explore large values of both $n$ and $N$, as well as various R\'enyi indices $\alpha$ and dimensionless interaction strengths $g$. We identify the numerical form of a general shape function $\Phi$ describing the scaling of the particle entanglement entropy for fermions 
\begin{equation}
S_\alpha = \ln{N\choose n}+N \Phi\qty(\sin \frac{n\pi}{N},N;g,\alpha)\, .
\label{Eq:motivatedscaling}
\end{equation}
We find:
\begin{enumerate}
\item $\Phi \sim \mathrm{O}(N^0)$ at fixed $n/N$ for $N \gg 1$; 
\item $\Phi \sim n/N$ for fixed $n$ at large $N$,  such that $S_\alpha \sim  n \qty[\ln N + \mathrm{O}(N^0)]$; 
\item $\Phi \to 0$ for $g\to0$ such that $S_\alpha = \ln {N \choose n}$ for non-interacting fermions. 
\end{enumerate}
Property (iii) can be understood as a direct result of the constant eigenvalues below the Fermi surface of the $n$-RDM for a single Slater determinant wavefunction describing non-interacting fermions \cite{Carlson:1961dr}.

The scaling of this shape function is confirmed by comparing with a bosonization calculation employing the known fermionic $1$-RDM (obtained from exponentials of bosonic correlation functions \cite{Giamarchi:2004bk}) for $\alpha = 2$, and a self-consistent non-linear least squares fitting analysis of numerical results up to $N=40$ fermions and $n \le 7$.

The main contributions of this work are: (1) an asymptotic expansion of the $1$-particle entanglement entropy in powers of $1/N$ which includes both integer and $g$-dependent exponents and high-precision estimates for multiplicative pre-factors for a 1D model of interacting fermions in the Luttinger liquid regime; (2) confirmation of the subleading extensive scaling (analogous to a volume law) in Eq.~\eqref{Eq:motivatedscaling} in the presence of interactions between fermions for general $n$ and $N$; (3) numerical determination of the scaling shape function $\Phi$, as a function of the ratio $n/N$; and (4) a procedure to obtain particle entanglement for large system sizes using limited data for small $n$ and $N$.  

The remainder of this paper is organized as follows. We begin with a detailed introduction to particle entanglement entropy and highlight previous results before defining the microscopic lattice model that is the focus of this work and describing its low energy sector.  Beginning with the analytically tractable case of $n=1$ and $\alpha=2$ we derive an asymptotic expansion of $S_2$, and discuss some implications for the case of general $n$. We then proceed to a careful analysis of a numerical data set for $S_2$ computed with exact diagonalization and the density matrix renormalization group. This allows for the confirmation of our bosonization procedure, a high precision non-linear fitting procedure to a general expansion form and ultimately a proposal for the shape function introduced in Eq.~\ref{Eq:motivatedscaling}.  The remainder of our results are presented in the context of understanding the form of $\Phi$ and we end with some remarks on implications for an improved understanding of particle entanglement and some future directions.

\section{Particle Entanglement Entropy}
\label{sec:particleEntanglement}

\begin{figure}
    \flushright
    \includegraphics[width=13.85cm]{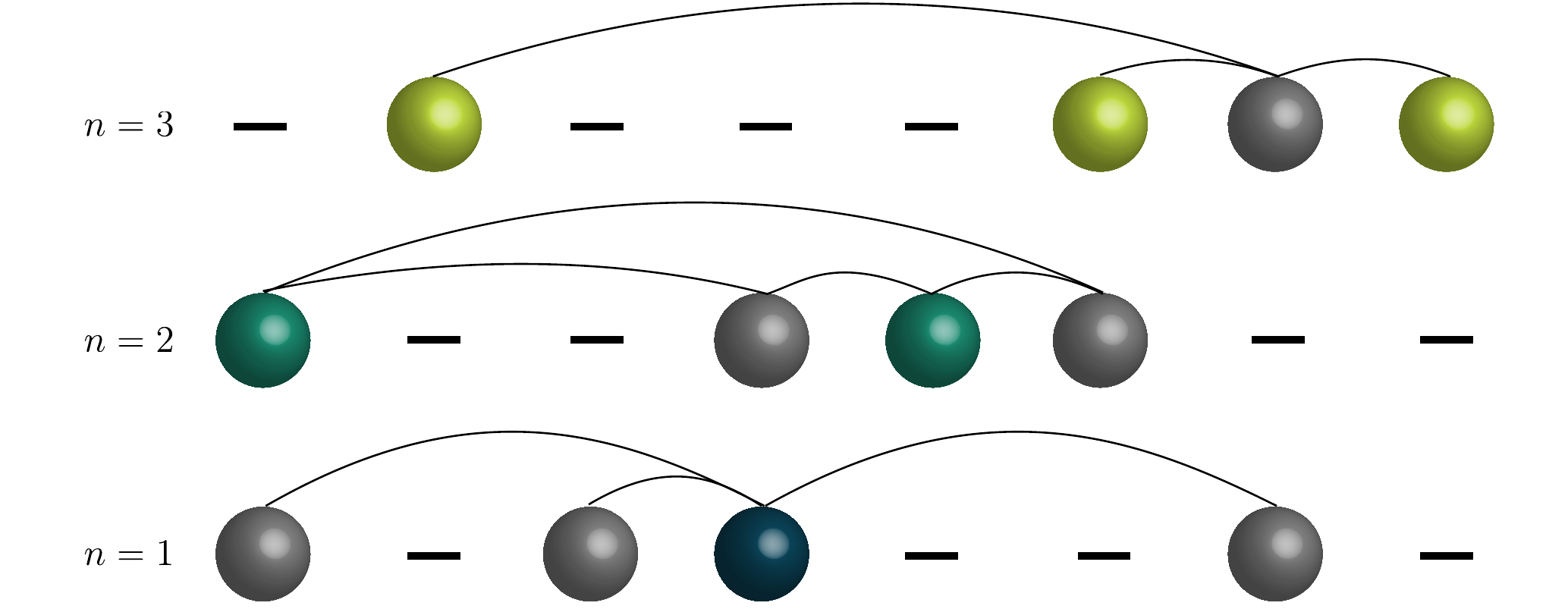}
    \caption{A quantum system of $N$ indistinguishable particles in one dimension can be partitioned into two subsets of $n$ and $N-n$ particles. Different values of little $n$ for $N=4$ are shown. The degrees of freedom traced out in the reduced density matrix are depicted in grey, while the colored ones are kept. Arcs indicate correlations between the two subsets. As seen for $n=1$ and $n=3$, the $n$-particle entanglement is equivalent to the $(N-n)$-particle entanglement.}
    \label{fig:01_particleent}
\end{figure}

A pure state $\ket{\Psi}$ describing $N$ indistinguishable particles is referred to as bipartite entangled when it cannot be expressed as a simple tensor product $\ket{\Psi} = \ket{\Psi_A} \otimes \ket{\Psi_B}$ with $\ket{\Psi_A} \in \mathcal{H}_A$, $\ket{\Psi_B} \in \mathcal{H}_B$ and $\mathcal{H}_A\otimes\mathcal{H}_B=\mathcal{H}$, where $\mathcal{H}$ is the Hilbert space containing $\ket{\Psi}$. The entanglement can be quantified by the \ren entanglement entropy, given by 
\begin{equation}
    S_\alpha(\rho_A) = \frac{1}{1-\alpha}\ln\left[\Tr(\rho_A^\alpha)\right] \ ,
\end{equation} 
where $\alpha$ is the \ren index and $\rho_A$ is the reduced density matrix of
the sub-Hilbert space $\mathcal{H}_A$ obtained by tracing out the degrees of
freedom in $\mathcal{H}_B$, i.e.,
$\rho_A=\Tr_B(\rho)=\Tr_B(\ket{\Psi}\bra{\Psi})$.  The familiar von Neumann
entropy is found from 
\begin{equation}
    S_1 (\rho_A) = -\Tr \rho_A \ln \rho_A = \lim_{\alpha \to 1} S_\alpha(\rho_A).
\end{equation}
Often, $A$ and $B$ correspond to partitions of $\mathcal{H}$ distinguished by observable modes of the system, with the most commonly studied  case being the division into non-overlapping spatial sub-regions. As discussed in the introduction, an alternate type of bipartition which is independent of the choice of modes and exploits the indistinguishiability of quantum particles involves decomposition of $\ket{\Psi}$ into two groups of particles \cite{Haque:2009zi} as seen in Fig.~\ref{fig:01_particleent}.  This is most natural to understand when the wavefunction is expressed in a first quantized form, and the resulting \emph{particle} entanglement entropy is then computed from $\rho_{A\equiv n}$ which is given by the familiar $n$-body reduced density matrix ($n$-RDM) \cite{Lowdin:1955qr,Ando:1963ic,coleman1963structure,sasaki1965eigenvalues} with an information theoretical normalization: $\Tr(\rho_n)=1$ (where $n$ is the number of particles in $A$). 
For a $N$ particle system confined on a 1D lattice, this can be computed by fixing $n$ coordinates $\qty{i_1,\dots,i_n}$ in the symmetrized wavefunction $\Psi(i_1,\dots,i_N) = \braket{i_1,\dots,i_N}{\Psi}$ and tracing out the remaining $N-n$ coordinates \cite{DelMaestro:2021ja}:
\begin{equation}
    \begin{aligned}
        \rho_n^{i_1,\dots,i_n;j_1,\dots,j_n} &= \sum_{i_{n+1},\ldots,i_N}\langle \Psi|i_1,\ldots,i_n,i_{n+1},\ldots,i_N\rangle\langle j_1,\ldots,j_n,i_{n+1},\ldots,i_N|\Psi\rangle  \ ,
    \end{aligned}
        \label{eq:nRDMdef}
\end{equation}
where we have explicitly included the $2n$ indices on the $n$-RDM. In higher dimensions, (or in the spatial continuum), Eq.~\eqref{eq:nRDMdef} can be generalized through appropriate coordinate labels.   General properties of the von Neumann entropy of the $n$-RDM include:
\begin{enumerate}
    \item monotonicity:  $S_1(\rho_n) \le S_1(\rho_{n+1})$ for $1 \le n \le N/2-1$;
    \item reflection:  $S_1(\rho_n) = S_1(\rho_{N-n})$;
    \item concavity: $S_1(\rho_n) \ge \qty[S_1(\rho_{n+1})+S_1(\rho_{n-1})]/2$ for $1 \le n \le N-1$.
\end{enumerate}
These properties guarantee that $S_1(\rho_{\lfloor N/2 \rfloor})$ is maximal  where $\lfloor \dots \rfloor$ denotes the integer part.  Property (ii) can be proven for general \ren index $\alpha$. To simplify notation in what follows, we will drop the explicit dependence on the $n$-RDM and shorten $S_\alpha(\rho_n) \equiv S_\alpha(n)$. 

For non-interacting bosons, the ground state $\ket{\Psi}$ can be described by a single tensor product state, and thus $\rho_n$ is pure and the resulting particle entanglement vanishes. In contrast, for non-interacting fermions, $\ket{\Psi}$ can be written as a single Slater determinant such that $\rho_n$ has $\binom{N}{n}$ identical eigenvalues \cite{sasaki1965eigenvalues} and $S_\alpha(n) = \ln \binom{N}{n}$, independent of the R{\'e}nyi index $\alpha$. 

For fermions it can be proven \cite{Carlen:2016,lemm2017entropy} that $S_1(\rho_n) \ge \ln N$ and it has been conjectured \cite{Haque:2007mu,Zozulya:2007fe,Zozulya:2008bg,Katsura:2007ty} that the fermionic particle entropy scales as
 \begin{equation}
      S_\alpha(n) \big \vert_{\text{fermions}} = \ln{N \choose n}+a_\alpha(n)+\mathcal{O}\left(\frac{1}{N^{\gamma_\alpha(n)}}\right) \ ,
     \label{eq:Sempirical}  
 \end{equation}
where $a_\alpha(n)$ and $\gamma_\alpha(n) \ge 1$ depend on both interactions and the number of particles in the subsystem. As mentioned above, for free fermions $a_{\alpha}(n) = 0$, and there are no $N$ dependent corrections as their prefactors vanish in the limit $V/J \rightarrow 0$.  Some other cases where the constant can be calculated include 1D charged density wave states where $a_{\alpha}(n)=\ln{2}$ \cite{Barghathi:2017tg}, and fermionic Laughlin states with $a_1(n)=-n\ln{\nu}$ where $\nu$ is the filling fraction \cite{Haque:2009zi}. 

Eq.~\eqref{eq:Sempirical} was verified for fermionic Luttinger liquids in Refs.~\cite{Barghathi:2017tg,Thamm:2022} for $n=1$ and $\alpha=2$, where $\gamma_2(1) = K + K^{-1} -1$ with $K$ the Luttinger parameter. For the low energy sector of the Lieb-Liniger model of $\delta$-function interacting bosons, an alternative scaling form was identified in Ref.~\cite{Herdman:2015xa} for the restricted cases of $\alpha = 2$ and $n \le 2$
 \begin{equation}
 S_2(n) \big \vert_{\text{bosons}} = \frac{n}{K}\ln{N} + b_\alpha(n)+\mathcal{O}\left(\frac{1}{N^{1-K^{-1}}}\right) \, .
     \label{eq:SempiricalBosons}
 \end{equation}
 Comparing Eqs.~\eqref{eq:Sempirical} and \eqref{eq:SempiricalBosons} highlights the sensitivity of particle entanglement to interactions and particle statistics at leading order but also demonstrates the paucity of results beyond some special cases in 1D. Work on this subject has been historically rather limited, possibly due to the fact that generating analytical results in a first quantized framework is challenging, and that it is generally not possible to experimentally address a subset of $n$ indistinguishable particles, calling into question the utility of particle entanglement as a potential resource for quantum information processing \cite{Ghirardi:2004,Tichy:2011,Balachandran:2013}. However, protocols now exist for the transfer of particle entanglement to addressable modes \cite{Killoran:2014am}, motivating new activity in understanding the role of entanglement in characterizing correlations between identical particles in many-body systems. The first step is to gain further knowledge about the subleading constants and $N$ dependent corrections that appear in the general particle entanglement scaling forms above, and in particular how they may scale with $n$, $N$ and $\alpha$. In this paper we focus exclusively on the fermionic case, where the universal form of the leading $\ln \binom{N}{n}$ term simplifies the issue.

\section{Model}
\label{sec:model}
\subsection{$J$-$V$ Hamiltonian and phase diagram}
\begin{figure}[H]
    \centering\includegraphics{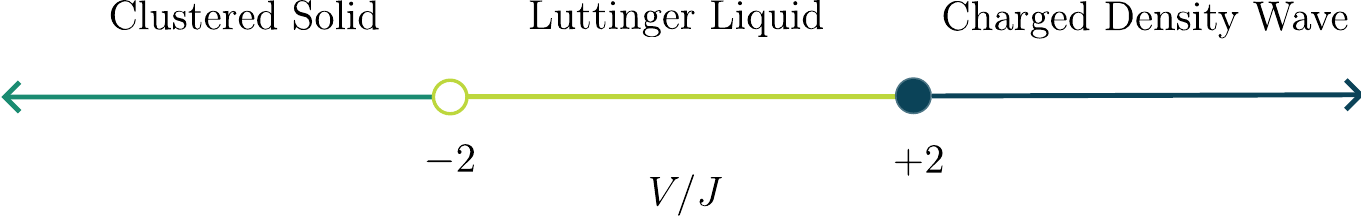}
    \caption{Phase diagram of the $J$-$V$ model. In the Tomonaga Luttinger liquid phase with $-2 < V/J < 2$, fermions delocalize. When $V/J \gg 1$ with repulsive interactions, fermions maximize their separation resulting in a charge-density wave (CDW) phase. The closed circle at $V/J=+2$ represents a continuous phase transition to the CDW phase. In the regime of attractive interactions $V/J \ll -2$, fermions cluster together resulting in a clustered solid phase. The open circle at $V/J=-2$ represents a first order phase transition.}
    \label{fig:01_phasediagram}
\end{figure}

We study the $J$-$V$ model of $N$ spinless fermions on a one dimensional lattice with $L$ sites at half-filling, $\rho_0=N/L=1/2$ (where lengths are measured in units of the lattice constant), which is described by the Hamiltonian \begin{equation}
    \begin{aligned}
    H &=-J\sum_{i=1}^L[c_{i+1}^\dagger c_i^{\phantom{\dagger}}+c_i^\dagger c_{i+1}^{\phantom{\dagger}}]+V\sum_{i=1}^Ln_in_{i+1} \ .
    \end{aligned}
    \label{Eq:jv_Hamiltonian}
\end{equation} 
Here, $c_i^\dagger (c_i)$ are the fermionic creation (annihilation) operators, and $n_i=c_i^\dagger c_i^{\phantom{\dagger}}$ is the occupation number operator for site $i$. The nearest neighbor hopping amplitude is given by $J>0$, and $V$ is the nearest neighbor interaction strength. To ensure the ground state is non-degenerate for an odd particle number $N$, we use periodic boundary conditions; while for even $N$, we use antiperiodic boundary conditions. By mapping Eq.~\eqref{Eq:jv_Hamiltonian} onto the XXZ spin-1/2 model, the phase boundaries can be determined as shown in Fig.~\ref{fig:01_phasediagram}, as it is exactly solvable via a Bethe ansatz \cite{DesCloizeaux:1966,Yang.1966}. 
At low energies and long wavelengths, the system can be described by Tomonaga-Luttinger liquid (TLL) theory when $-2 < V/J < 2$, and the corresponding TLL parameter $K$ is given by \cite{Haldane:1980} 
\begin{equation}
    \begin{aligned}
    K &=  \frac{\pi}{2\cos^{-1}{(-\frac{V}{2J})}} \ .
    \end{aligned}
    \label{Eq:ll_parameter}
\end{equation}
Attractive interactions ($V<0$) have a corresponding $K > 1$. When the
interaction strength is increased, the system undergoes a continuous phase
transition to the charge-density wave phase at $V=2,K=1/2$. Repulsive
interactions ($V > 0$) have $1/2 \leq K < 1$.  Decreasing $V/J$ results in a first order phase transition to the clustered solid phase at $V/J=-2,K\to\infty$. Free fermions $(V=0)$ correspond to TLL parameter $K=1$.

\subsection{Tomonaga-Luttinger liquid theory and bosonization}

In the Luttinger liquid regime, which is characterized by relatively weak
interactions and long wavelength fluctuations, the dynamics of the system are
primarily governed by low-energy excitations in the form of density variations
about a mean density \cite{Giamarchi:2004bk}. In this region, the microscopic
Hamiltonian in Eq.~\eqref{Eq:jv_Hamiltonian}, can be recast in terms of these
bosonic density variations, enabling us to write an effective Hamiltonian in bosonization language to account for the low-energy physics. Following the linearization of the energy dispersion near the Fermi points \cite{Haldane:1981eh,Iucci:2009lg}
\begin{equation*}
    \begin{aligned}
        H &= \sum_{q\neq 0}[\omega_0(q)+m(q)]b_q^\dagger b_q + \frac{1}{2}\sum_{q \neq 0}g_2(q)[b_qb_{-q}+b_q^\dagger b_{-q}^\dagger] \ ,
    \end{aligned}
\end{equation*}
where $b_q (b_q^\dagger)$ are bosonic annihilation (creation) operators,
$\omega_0(q)=v_F|q|$, and $q$ are discrete momenta: $q_n=2\pi n/L, n\in
\mathbb{Z}\setminus \{0\}$. For small $q$, $g_2(q)=g_2|q|$ and $m(q)=g_4|q|$ where $g_2$
and $g_4$ are parameters obtained from the mapping from the $J$-$V$ Hamiltonian. To account for the short-range interactions in the $J$-$V$ model, an interaction cutoff $\epsilon$ is implemented such that $m(q)$ and $g_2(q)$ vanish for $q\epsilon \gg 1$.
This Hamiltonian is quadratic with respect to the boson operators, in contrast
to the fermionic Hamiltonian which is of fourth order. Utilizing
bosonization, we recast operators for the fermionic field in terms
of exponentiated bosonic fields and thus calculate the one-point
correlation function for both right- and left-movers, yielding the one-body density matrix. 
\begin{equation*}
    \begin{aligned}
        C_{\pm}(x) &= \langle\Psi_{\pm}^\dagger(x)\Psi^{\phantom\dagger}_{\pm}(0)\rangle \\
        \rho_1(x) &=\frac{1}{N}[e^{-ik_Fx}C_+(x)+e^{ik_Fx}C_-(x)] \ ,
    \end{aligned}
\end{equation*}
where $\Psi_\alpha$ are the fermionic field operators, $\pm$ corresponds to right/left movers, and $k_F = N\pi/L$ is the Fermi momentum. For full details on the analytic computation of the one body reduced density matrix in this model within the bosonization framework, see Ref.~\cite{Thamm:2022}.

\section{1-Particle entanglement in fermionic Tomonaga-Luttinger liquids}
\label{sec:onePartCalc}

We begin by analytically deriving the asymptoptic finite size scaling for the R\'{e}nyi  entanglement entropy with $\alpha=2$ for a bipartition of $n=1$ and $N-1$ particles which is given by 
\begin{equation}
    \begin{aligned}
    S_2(n=1)=-\ln \mathrm{Tr}\rho_1^2 \ ,
    \end{aligned}
\end{equation} 
where $\rho_1$ is the one body reduced density matrix at zero temperature. Taking into account periodic boundary conditions, the distance, or chord length, between two points is $\frac{L}{\pi}\sin{\frac{\pi}{L}|x_1-x_2|}$. 

\subsection{Non-interacting spinless fermions}

Starting with the non-interacting case ($V=0$, indicated by a superscript 0), the one body density matrix is given by 
\begin{equation}
    \begin{aligned}
        \rho_{1}^0 &= \frac{1}{N}\frac{\sin{(k_F|x_1-x_2|)}}{L\sin{(\pi|x_1-x_2|/L)}} \ ,
    \end{aligned}
    \label{Eq:ff_obdm}
\end{equation}
where $k_F=\frac{N\pi}{L}$. To compute the entanglement entropy we need the trace of this quantity squared, 
\begin{equation}
    \begin{aligned}
        \Tr[(\rho_1^0)^2]&= \int_{-L/2}^{L/2}dx_2\int_{-L/2}^{L/2}dx_1\frac{1}{N^2}\frac{\sin^2{(\pi\rho_0|x_1-x_2|)}}{L^2\sin^2{(\pi|x_1-x_2|/L)}} \\
        &= \frac{2}{N^3}\int_0^{N/2}dy\frac{\sin^2{(\pi y)}}{\sin^2(\pi y/N)} \ ,
    \end{aligned}
    \label{Eq:trace_ff_cl}
\end{equation}
where we have used the translational invariance of the system and defined
$y=\rho_0|x_1-x_2|$. While this is analytically solvable in the case of free
fermions (and gives $1/N$), for interacting fermions we will later require the
use of some approximations (ignoring
the effects of periodic boundary conditions and replacing the chord length with separations
$\frac{L}{\pi}\sin{\frac{\pi}{L}|x_1-x_2|}\rightarrow |x_1-x_2|$) whose
accuracy can be demonstrated here.  Replacing the chord length in
the denominator of Eq.~\eqref{Eq:ff_obdm} yields a modified density matrix
(indicated by a tilde) that is correct only for separations $\abs{x_1-x_2} \ll L$: 
\begin{equation}
    \begin{aligned}
        \rhonocl^0  &= \frac{\sin{(\pi\rho_0|x_1-x_2|)}}{\pi\rho_0L|x_1-x_2|}  \ . 
    \end{aligned}
\end{equation}
The trace of the square of the modified density matrix is then given by
\begin{equation}
       \begin{aligned}
           \Tr[(\rhonocl^0)^2] &= \int_{-L/2}^{L/2}dx_2\int_{-L/2}^{L/2}dx_1\frac{\sin^2{(\pi\rho_0|x_1-x_2|)}}{\pi^2\rho_0^2L^2|x_1-x_2|^2} \\
           &= \frac{2}{N}\int_{0}^{N/2}dy\frac{\sin^2{(\pi y)}}{\pi^2y^2} \ .
       \end{aligned}
       \label{eq:Trrho02}
   \end{equation}
We can rewrite Eq.~\eqref{Eq:trace_ff_cl} by adding and subtracting
Eq.~\eqref{eq:Trrho02} as 
\begin{equation}
   \begin{aligned}
       \Tr[(\rho_1^0)^2] &= \frac{2}{N^3}\int_0^{N/2}dy\frac{\sin^2{(\pi y)}}{\sin^2(\pi y/N)}-\frac{2}{N}\int_{0}^{N/2}dy\frac{\sin^2{(\pi y)}}{\pi^2y^2}+\frac{2}{N}\int_{0}^{N/2}dy\frac{\sin^2{(\pi y)}}{\pi^2y^2} \\
       &= f(y,N)+g(y,N) 
   \end{aligned}
\end{equation}
where:
\begin{align}
\label{eq:fyN}
    f(y,N) &= \frac{2}{N^3}\int_0^{N/2}dy\frac{\sin^2{(\pi y)}}{\sin^2(\pi
    y/N)}-\frac{2}{N}\int_{0}^{N/2}dy\frac{\sin^2{(\pi y)}}{\pi^2y^2} \\ 
\label{eq:gyN}
        g(y,N)&= \frac{2}{N}\int_{0}^{N/2}dy\frac{\sin^2{(\pi y)}}{\pi^2y^2} \, .
\end{align}
To compute $f(y,N)$, we use that the integral of $\sin^2(x)$ is $1/2$ over its
period $2\pi$, and thus we approximate the highly oscillating function $\sin^2{(\pi y)}\rightarrow \frac{1}{2}$. In the large $N$ limit this gives 
\begin{equation}
    \begin{aligned}
        f(y,N) &= \frac{1}{N}\left[\frac{1}{N^2}\int_0^{N/2}dy\frac{1}{\sin^2{(\pi y/N)}}-\int_0^{N/2}dy\frac{1}{\pi^2y^2}\right] \\
        &= \frac{2}{N^2\pi^2} \ .
    \end{aligned}
    \label{Eq:f_ff}
\end{equation}
While we can compute $g(y,N)$ directly, we will demonstrate another approximation that is necessary for the interacting case. We rewrite
\begin{equation}
    \begin{aligned}
        g(y,N)&= \frac{2}{N}\int_{0}^{N/2}dy\frac{\sin^2{(\pi y)}}{\pi^2y^2} \\
        &= \frac{1}{N}\left[\int_{0}^{\infty}dy\frac{2\sin^2{(\pi y)}}{\pi^2y^2}-\int_{N/2}^{\infty}dy\frac{2\sin^2{(\pi y)}}{\pi^2y^2}\right] \ .
    \end{aligned}
\end{equation}
The first term can be directly integrated and in the second term, we again
approximate $\sin^2{(\pi y)}\rightarrow 1/2$.  This gives 
\begin{equation}
     \begin{aligned}
         g(y,N) &= \frac{1}{N}\left[\int_{0}^{\infty}dy\frac{2\sin^2{(\pi y)}}{\pi^2y^2}-\int_{N/2}^{\infty}dy\frac{1}{\pi^2y^2}\right] \\
         &= \frac{1}{N}\left[1-\frac{2}{N\pi^2}\right] \ .
     \end{aligned}
     \label{Eq:g_ff}
 \end{equation}
Combining, Eq.~\eqref{Eq:f_ff} and \eqref{Eq:g_ff} yields
\begin{equation}
\begin{aligned}
    \Tr[(\rho_1^0)^2] &=\frac{2}{N^2\pi^2} + \frac{1}{N}\left[1-\frac{2}{N\pi^2}\right]\\
    &= \frac{1}{N}\\
    \rightarrow S_2(1)& = -\ln{\Tr[(\rho_1^0)^2]} = \ln{N} \,
    \end{aligned}
    \label{Eq:ent_ff}
\end{equation}
which is the expected result for free fermions (for a full derivation of the
exact result see \ref{sec:ffcalc}), supporting the use of
the replacement of $\sin^2{\pi y}$ for $N$ large. 

\subsection{1-Particle entanglement in Luttinger Liquid}
\begin{figure}[H]
    \centering
    \includegraphics{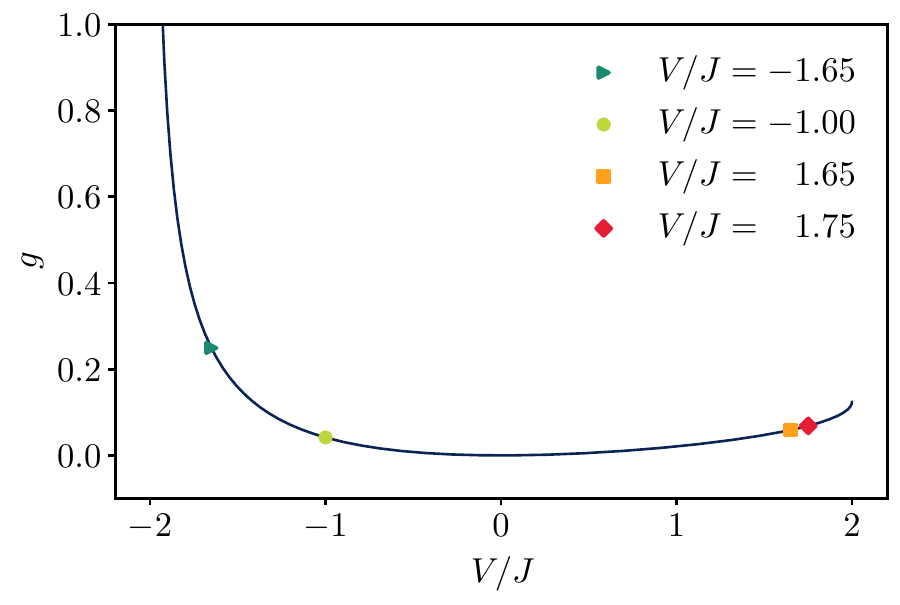}
    \caption{Values of the exponent $g$ in the one body density matrix as a
    function of interaction strength $V/J$ in the Tomonaga-Luttinger Liquid
phase. Markers correspond to the $g$ values of the interaction strengths we
primarily consider.}
    \label{fig:03_g_interaction}
\end{figure}

 We now consider interacting fermions. At zero temperature in the thermodynamic limit, the reduced density matrix for $n=1$ is given by  \cite{Thamm:2022,Cazalilla:2006zw}
\begin{equation}
    \begin{aligned}
       \rho_1(x_1,x_2) &= \frac{1}{N}\frac{\sin{(k_F|x_1-x_2|)}}{L\sin{(\pi |x_1-x_2|/L)}}\left|\frac{\sin{(\pi i\epsilon/L)}}{\sin{(\frac{\pi}{L}(|x_1-x_2|+i\epsilon))}}\right|^{2g} \\
       &= \frac{1}{N}\frac{\sin{(k_F|x_1-x_2|)}}{L\sin{(\pi |x_1-x_2|/L)}}\left|\frac{1}{1+\sin^2{
       (\pi |x_1-x_2|/L)}/\sinh^2{(\pi\epsilon/L)}}\right|^{g} \ ,
    \end{aligned}
    \label{Eq:rho_CL} 
\end{equation}
where $\epsilon$ is an interaction dependent short distance cutoff, and $g=
(K+K^{-1}-2)/4$. The values of $g$ for different interactions are shown in Fig.~\ref{fig:03_g_interaction}. 
Squaring and computing the trace:
\begin{align}
    \Tr[\rho_1^2] &=
    \int_{-L/2}^{L/2}dx_2\int_{-L/2}^{L/2}dx_1\frac{\sin^2{(k_F|x_1-x_2|)}}{N^2L^2\sin^2{(\pi
    |x_1-x_2|/L)}}\frac{1}{\left[1+\frac{N^2}{\pi^2\epsilon^2\rho_0^2}\sin^2{(\pi
    |x_1-x_2|/L)}\right]^{2g}}\nonumber \\  
&=\frac{2}{N^3}\int_0^{N/2}dy\frac{\sin^2{\pi y}}{\sin^2{(\pi y/N)}}\frac{1}{\left[\frac{N^2}{\pi^2\rho_0^2\epsilon^2}\sin^2{(\pi y/N)}+1\right]^{2g}} \ ,
    \label{Eq:ent_cl_integral}
\end{align}
where we have made the same change of variables as before, $y=\rho_0|x-x'|$. We have further  approximated $\sinh{\frac{\pi\epsilon}{L}}\approx\frac{\pi\epsilon}{L}$. Though this quantity cannot be integrated directly, we can utilize the same technique that we employed for free fermions,
\begin{equation}
    \begin{aligned}
    \Tr[\rho_1^2] &= \frac{1}{N}
    \Bigg[\frac{1}{N^2}\int_0^{N/2}dy\frac{2\sin^2{\pi y}}{\sin^2{\pi y/N}}\frac{1}{\left[\frac{N^2}{\pi^2\rho_0^2\epsilon^2}\sin^2{(\pi y/N)}+1\right]^{2g}}
    \\
    &\ \ \ \ \ \ \ \ \ \
    -\int_{0}^{N/2}dy\frac{2\sin^2{(\pi y)}}{\pi^2y^2[1+y^2/\rho_0^2\epsilon^2]^{2g}}
    \Bigg] + \frac{1}{N}\int_{0}^{N/2}dy\frac{2\sin^2{(\pi y)}}{\pi^2y^2[1+y^2/\rho_0^2\epsilon^2]^{2g}} \ .
    \end{aligned}
    \label{Eq:ent_withcl}
\end{equation}
Evaluating the first two terms by replacing $\sin^2{(\pi y)}\rightarrow
\frac{1}{2}$ and in the large $N$ limit yields,  
\begin{equation}
    \begin{aligned}
          \Delta &= \frac{1}{N}\left[\frac{2}{N\pi^2}\; \hypergeometric\left(-\frac{1}{2};2g,\frac{1}{2};-\frac{N^2}{4\epsilon^2\rho_0^2}\right)-\frac{gN\hypergeometric\left(\frac{1}{2};1+2g,2;-\frac{N^2}{\pi^2\epsilon^2\rho_0^2}\right)}{\pi^2\epsilon^2\rho_0^2}\right] \ , 
    \end{aligned}
    \label{Eq:diff_cl_andnocl}
\end{equation}
where $\hypergeometric(a_1;b_1,b_2,z)$ is the generalized hypergeometric function. 
The term added in Eq.~\eqref{Eq:ent_withcl} is given by 
\begin{multline}
      \frac{1}{N}\int_{0}^{N/2}dy\frac{2\sin^2{(\pi y)}}{\pi^2y^2(1+y^2/\epsilon^2\rho_0^2)^{2g}} 
       \\
       = \frac{1}{N}\int_{0}^{\infty}dy\frac{2\sin^2{(\pi
       y)}}{\pi^2y^2(1+y^2/\rho_0^2\epsilon^2)^{2g}} - \frac{1}{N}\int_{N/2}^{\infty}dy\frac{2\sin^2{(\pi y)}}{\pi^2y^2(1+y^2/\rho_0^2\epsilon^2)^{2g}} \ .
    \label{Eq:ent_noCL} 
\end{multline}

The first integral can be computed exactly: 
\begin{equation}
    \begin{aligned}
      \frac{1}{N}\int_{0}^{\infty}dy\frac{2\sin^2{(\pi y)}}{\pi^2y^2(1+y^2/\rho_0^2\epsilon^2)^{2g}} &= \frac{\rho_0^{4g}\epsilon_{\phantom{0}}^{4g}\pi_{\phantom{0}}^{\frac{1}{2}+4g}\hypergeometric\left(2g; 1+2g,\frac{3}{2}+2g;\pi^2\rho_0^2\epsilon^2\right)\sec{(2g\pi)}}{2N\Gamma(1+2g)\Gamma(\frac{3}{2}+2g)} \\
      &+\frac{\Gamma(\frac{1}{2}+2g)}{N\epsilon\rho_0\pi^{\frac{3}{2}}\Gamma(2g)}\left[\hypergeometric\left(-\frac{1}{2};\frac{1}{2},\frac{1}{2}-2g,\pi^2\epsilon^{2}\rho_0^2\right)-1\right] \ ,
    \label{Eq:t1_noCL} 
    \end{aligned}
\end{equation}
where $\Gamma(z)$ is the Euler gamma function. For the second integral, we again send $\sin^2(\pi y)\rightarrow 1/2$, so the denominator can be integrated in the large $N$ limit.
\begin{equation}
    \begin{aligned}
      \frac{1}{N}\int_{N/2}^{\infty}dy\frac{1}{\pi^2y^2(1+y^2/\rho_0^2\epsilon^2)^{2g}} &= -\frac{i(-1)^{-2g}\beta\left(-\frac{4\epsilon^2\rho_0^2}{N^2},\frac {1}{2}+2g,1-2g\right)}{2\epsilon\rho_0\pi^2} \ ,
    \end{aligned}
    \label{Eq:t2_noCL} 
\end{equation}
where $\beta(p,q)$ is the beta integral.  To obtain the final expression for the finite-size corrected quantity Eq.~\eqref{Eq:ent_withcl}, we plug in the terms derived in Eq.~\eqref{Eq:diff_cl_andnocl}, Eq.~\eqref{Eq:t1_noCL}, and Eq.~\eqref{Eq:t2_noCL} 
\begin{equation}
    \begin{aligned}
    \Tr[\rho_1^2]&= \frac{1}{N}\left[\frac{2}{N\pi}\;\hypergeometric\left(-\frac{1}{2};2g,\frac{1}{2};-\frac{N^2}{4\epsilon^2\rho_0^2}\right)-\frac{gN\hypergeometric\left(\frac{1}{2};1+2g,2;-\frac{N^2}{\pi^2\epsilon^2\rho_0^2}\right)}{\pi^2\epsilon^2\rho_0^2}\right] \\
& \quad+ \frac{\rho_0^{4g}\epsilon_{\phantom{0}}^{4g}\pi_{\phantom{0}}^{\frac{1}{2}+4g}\hypergeometric\left(2g; 1+2g,\frac{3}{2}+2g;\pi^2\rho_0^2\epsilon^2\right)\sec{(2g\pi)}}{2N\Gamma(1+2g)\Gamma(\frac{3}{2}+2g)} \\
      &\quad
    +\frac{\Gamma(\frac{1}{2}+2g)}{N\epsilon\rho_0\pi^{\frac{3}{2}}\Gamma(2g)}\left[\hypergeometric\left(-\frac{1}{2};\frac{1}{2},\frac{1}{2}-2g,\pi^2\epsilon^{2}\rho_0^2\right)-1\right]
   \\ &\quad + \frac{i(-1)^{-2g}\beta\left(-\frac{4\epsilon^2\rho_0^2}{N^2},\frac {1}{2}+2g,1-2g\right)}{2\epsilon\rho_0\pi^2} \ .
    \end{aligned}
    \label{Eq:ent_cl_full}
\end{equation}
Eq.~\eqref{Eq:ent_cl_full} represents a new result that can be checked by
taking the limit $g\rightarrow 0$ $(K=1)$. In this case, $\Tr \rho_1^2 = 1/N$,
and thus the second \ren entropy at $n=1$ is given by $\ln{N}$,
the exact result for free fermions on a lattice \cite{Zozulya:2008bg,Barghathi:2017tg}. 

Using Eq.~\eqref{Eq:ent_cl_full}, the asymptotic scaling form for the second
\ren entropy at $n=1$ can be obtained by dividing out $1/N$ and expanding:
\begin{equation}
    \begin{aligned}
    e^{-\Delta S_2(n=1)}=N\Tr[\rho_1^2] &= \frac{\Gamma(\frac{1}{2}+2g)[\hypergeometric\left(-\frac{1}{2};\frac{1}{2},\frac{1}{2}-2g;\pi^2\epsilon^2\rho_0^2\right)-1]}{\pi^{3/2}\epsilon\rho_0\Gamma(2g)}\\
& \quad +\frac{(2\pi)^{4g}(\epsilon\rho_0)^{4g}\hypergeometric\left(2g;1+2g,\frac{3}{2}+2g;\pi^2\epsilon^2\rho_0^2\right)\sec{(2g\pi)}}{\Gamma(2+4g)} \\
&\quad + \frac{\sqrt{\pi}\epsilon\rho_0\Gamma(-\frac{1}{2}+2g)}{4\Gamma(2g)N^2}+\frac{3\pi^{5/2}(\epsilon\rho_0)^3\Gamma(-\frac{3}{2}+2g)}{32\Gamma(2g)N^4} \\
&\quad - \frac{2g\pi^{-\frac{1}{2}+4g}(\epsilon\rho_0)^{4g}\Gamma(-\frac{1}{2}-2g)}{2\Gamma(1-2g)}\frac{1}{N^{1+4g}}\\
&\quad -\frac{2g\pi^{\frac{3}{2}+4g}(\epsilon\rho_0)^{2+4g}\Gamma(-\frac{3}{2}-2g)}{2\Gamma(-1-2g)}\frac{1}{N^{3+4g}}\\
&\quad +\mathcal{O}\left(\frac{1}{N}\right)^{{\text{min}}\{6,5+4g\}} \ ,
    \end{aligned}
    \label{Eq:ent_cl_expanded}
\end{equation}
where we have introduced the notation: 
\begin{equation}
    \Delta S_\alpha(n) \equiv S_\alpha(n)-\ln\binom{N}{n}\, .
\label{eq:DeltaSDef}
\end{equation}
Defining $C_0(\alpha=2,n=1)$ as the $N$-independent term in Eq.~\eqref{Eq:ent_cl_expanded}, we can write
 \begin{equation}
      \Delta S_2(n=1) = -\ln C_0(\alpha=2,n=1) +\mathcal{O}\left(\frac{1}{N^{\min\{2,1+4g\}}}\right) 
     \label{eq:Sn1alpha2}
 \end{equation}
which is consistent with conjectured scaling of fermionic particle entanglement entropy Eq.~\eqref{eq:Sempirical}.
\subsection{n-Particle entanglement}

Without information about the form of the higher $n$ density matrices, it is
difficult to determine an exact form for the \ren entropy. However, the
general properties of the entanglement discussed in
Section~\ref{sec:particleEntanglement} can be used to direct a systematic
analysis of numerical data to construct a shape function for the general
$n$-particle entanglement entropy. First, the result for free fermions is given
by $\ln \binom{N}{n}$ and thus the remaining terms must vanish in the absence
of interactions suggesting that the coefficients of each subleading term are
dependent on interaction strength $g$. Furthermore, this leading order
scaling is independent of the \ren index, so the $\alpha$ dependence can only
appear in the subleading terms, either in the coefficients or as exponents of
finite size corrections. Finally, we know that the $n$-particle entanglement is
equivalent to the $N-n$-particle entanglement, suggesting the entanglement as a
function of $n/N$ reaches a maximum at $n=\lfloor N/2 \rfloor $ and is symmetric about this point.

\section{Results}
\label{sec:results}

To empirically construct the shape function describing the $n$-particle entanglement
entropy for the ground state of the $J$-$V$ Hamiltonian given by
Eq.~\eqref{Eq:jv_Hamiltonian}, we carry out large scale numerical computations
using both exact diagonalization (ED) and approximate density matrix
renormalization group (DMRG) techniques. While ED gives approximation-free
results, the exponential dependence of the size of the Hilbert space on the
length of the system $L$ limits the possible values of $(n,N,L)$ that we can
consider. To mitigate this, we encode the basis states with an integer fermion
basis and exploit the translational, inversion, and particle-hole symmetries of
the system, increasing the efficiency of operations performed on the states and
reducing the size of the space we need to consider by up to a factor of
$1/(4L)$. This allows us to perform exact diagonalization for even the
$7$-particle entanglement at $L=28$ sites. We further supplement the ED data
with DMRG results to study much larger systems at the cost of truncation
errors, using ED to benchmark the data at small system sizes. Utilizing the
\texttt{ITensors.jl} \cite{Itensor} package, the ground state is obtained as a
matrix product state which is then used to compute the $n$-RDM and, by
extension, the $n$-particle entanglement entropy. To reduce the necessary
computation, several tricks are deployed such as mapping the fermion operators
onto an integer basis, and leveraging the anticommutation relations of the fermion operators. This allows us to compute the two-particle entanglement entropy for systems up to $L=80$ lattice sites, the three-particle entanglement entropy for up to $L=48$, and even the four-particle entanglement entropy for up to $L=32$ sites. 
\footnote{While it is possible to compute higher order reduced density matrices
with DMRG we find that systematic errors related to the number of states we
keep begin to affect the ability to perform high precision non-linear fits to
proposed scaling functions.} 

Further detail on the implementation of ED and DMRG is provided in
\ref{app:methodology}.

All code to reproduce the numerical simulations, fits, and figures is open
source and has been made available online \cite{CODEHarini}.


\subsection{1-Particle entanglement}
\begin{figure}[H]
    \flushleft\hspace{2.5cm}
    \includegraphics{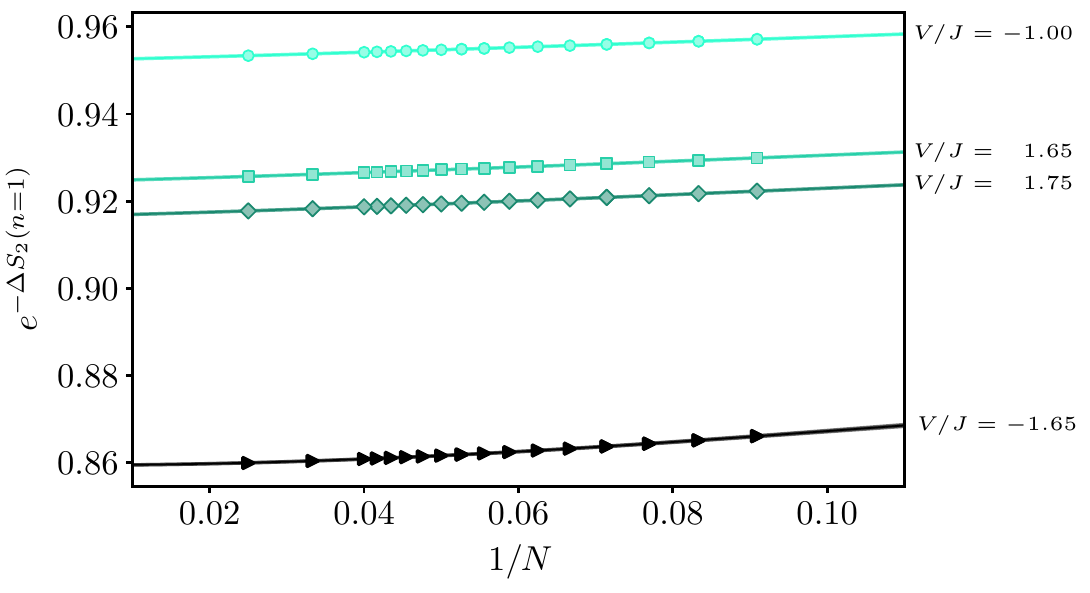}
    \caption{$e^{-\Delta S_2(n=1)}$ for different $V/J$ at $\alpha=2$. Solid lines
    correspond to fitting the data according to
    Eq.~\eqref{Eq:fittingform}. Symbols correspond to data for $N>10$ collected
from DMRG and ED at four different values of the interaction strength $V/J$
indicated by the different markers and colors.}
    \label{fig:04_alpha2n1}
\end{figure}

In this section we exclusively study the 1-particle entanglement ($n=1$).
Using ED and DMRG, we numerically test our analytic scaling at
$\alpha=2$ and $n=1$ to (i) verify finite size corrections in powers of $1/N$
and (ii) determine the coefficients of the scaling along with their error and
relative importance.  This procedure will serve as a benchmark for 
determining how the scaling of the shape function depends on $\alpha$.
Finally, we will motivate generalizing to $n>1$ where no analytic form for the
particle entanglement entropy is known to exist. To simplify our analysis, we
focus on the exponentiated subleading terms in the particle entanglement
which admit the expansion (using as yet unknown coefficients $C_j$)
\begin{equation}
\begin{aligned}
    e^{-\Delta S_2(n=1)} &= C_0+\frac{C_{1+4g}}{N^{1+4g}}+\frac{C_2}{N^{2}}+\frac{C_{3+4g}}{N^{3+4g}}+\frac{C_4}{N^{4}}\ ,
    \end{aligned} \label{Eq:fittingform}
\end{equation}

where we only include terms up to fourth order in $1/N$ as was done previously
in Eq.~\eqref{Eq:ent_cl_expanded}.  To test this scaling, we calculate
$S_2(n=1)$ deep within the TLL phase, away from the phase transitions, using the
ground state of the $J$-$V$ model. At each interaction strength, exponents are
computed using $g=(K+K^{-1}-2)/4$ with $K$ determined by
Eq.~\eqref{Eq:ll_parameter}. The coefficients of Eq.~\eqref{Eq:fittingform} are
extracted from a polynomial fit to $e^{-\Delta S_2(n=1)}$. The results are
illustrated in Fig.~\ref{fig:04_alpha2n1} where we include only system sizes with $N > 10$ along with a fit to the form given by Eq.~\eqref{Eq:fittingform}. 
 
The markers show $e^{-(S_2(n=1)-\ln{N})}$ results from ED and DMRG, while the
solid lines correspond to the polynomial fit,  and the data exhibits excellent agreement with the predicted scaling form. 
The numerical values of the coefficients of the $N$ dependent terms and the
constant term $C_0$ in Eq.~\eqref{Eq:fittingform} are presented in
Table~\ref{tab:01_coeffsalpha2} and Table~\ref{tab:02_constantalpha2}
respectively. We utilize parenthetical notation for the error to denote
modification of the last digit; for example, $C_4$ at $V/J=-1.65$ is expressed
as $-13.5(7) \equiv -13.5\pm 0.7$. We additionally report the constant terms predicted by our bosonization formula Eq.~\eqref{Eq:ent_cl_expanded}%
\begin{align}
   C_0(g) &= \frac{\Gamma(\frac{1}{2}+2g)[\hypergeometric\left(-\frac{1}{2};\frac{1}{2},\frac{1}{2}-2g;\pi^2\epsilon^2\rho_0^2\right)-1]}{\pi^{3/2}\epsilon\rho_0\Gamma(2g)}\nonumber \\
& \quad +\frac{(2\pi)^{4g}(\epsilon\rho_0)^{4g}\hypergeometric\left(2g;1+2g,\frac{3}{2}+2g;\pi^2\epsilon^2\rho_0^2\right)\sec{(2g\pi)}}{\Gamma(2+4g)}\, .
\label{eq:C0galpha2}
\end{align}
To facilitate comparison, the non-universal cutoff $\epsilon$ has been extracted using the same approach presented in Ref.~\cite{Thamm:2022}. We find close agreement between the predicted values for $C_0$ and the extracted values from the fit.

\begin{table} [H]
\caption{Interaction dependence for $\alpha = 2$. Coefficients for Eq.~\eqref{Eq:fittingform} at different values of $V/J$. For the special interaction strength $V/J = -1.65 \simeq -2\cos[\pi/(3+\sqrt{5})]$, $g\simeq 1/4$ and thus we fit to a reduced form containing only even inverse powers of $1/N$ corresponding to $C_2$ and $C_4$.}
\centering
\begin{tabular}{rcccc}
\multicolumn{5}{c}{$n=1,\ \alpha = 2$} \\
\toprule
{$V/J$}&{$C_{1+4g}$}&{$C_2$}&{$C_{3+4g}$}&{$C_4$} \\ 
\midrule
$-1.65$ & $-$& $\quad\!0.918(6)$ & $-$& $-13.5(7)$\\
$-1.00$ & $8.1197(1)\times 10^{-2}$ & $-1.070(1)\times 10^{-2}$& $1.3(2)\times 10^{-3}$& $-1.58(9)\times 10^{-2}$\\
$1.65$ & $0.11576(1)$& $-7.52(1)\times 10^{-2}$ & $0.176(3)$& $-0.31(1)$\\
$1.75$& $0.13827(2)$& $-1.078(3)\times 10^{-2}$& $0.368(8)$ & $-0.69(2)$\\
\bottomrule
\end{tabular}
\label{tab:01_coeffsalpha2}
\end{table}

We can extract the leading order terms, (constants in the scaling form), with
high precision, determining uncertainties on the order of $10^{-5}$ for
$V/J=-1.65$ and $10^{-7}$ otherwise. The increase in error for $V/J=-1.65$ is
expected due to proximity to the first order phase transition at $V/J=-2$ where
the TLL picture should begin to break down. 
\begin{table}[H]
    \centering
        \caption{Interaction dependence for $\alpha = 2$. Leading order coefficient/constant term in Eq.~\eqref{Eq:fittingform} at different values of $V/J$.}
\begin{tabular}{rll} 
\multicolumn{2}{c}{$n=1,\ \alpha = 2$} \\
    \toprule
     {$V/J$}& {$C_0|_{\rm predicted}$} & {$C_0|_{\rm fit}$}\\ \midrule
     $-1.65$ & $0.85555$&$0.85945(1)$\\
     $-1.00$ & $0.95170138$&$0.95229438(1)$  \\
     $1.65$ &$0.92378681$&$0.92452681(8)$\\
     $1.75$ & $0.915705201$&$0.91659419(9)$ \\
     \bottomrule
\end{tabular}
    \label{tab:02_constantalpha2}
\end{table}
We now attempt to identify the dependence of the asymptotic scaling of the shape function on the
order of the \ren entanglement entropy $\alpha$. 

\begin{figure}[H]
    \centering
    \includegraphics{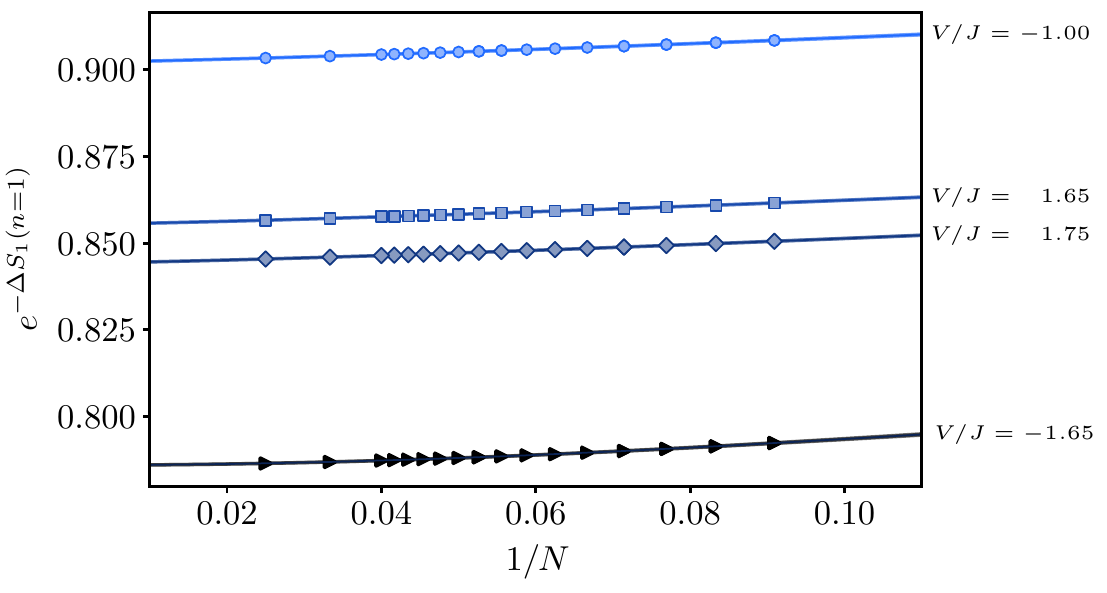}
    \caption{$e^{-\Delta S_1(n=1)}$ for different interaction strengths $V/J$
    and fixed $\alpha=1$. Solid lines correspond to fitting to
\eqref{Eq:fittingform} while points correspond to data collected from DMRG and
ED with different markers used to indicate four values of $V/J$. }
    \label{fig:05_alpha1n1}
\end{figure}

For $\alpha=1$ (the von Neumann case), we extract the coefficients in Eq.~\eqref{Eq:fittingform} through a polynomial fit. The results are shown in Fig.~\ref{fig:05_alpha1n1} for the same values of $V/J$ used to test $\alpha=2$.

The numerical values of the coefficients extracted from the hypothesized
scaling relation and the leading order terms are shown in Tables
\ref{tab:03_coeffsalpha1} and \ref{tab:04_constantalpha1} respectively. We find
that the errors in the coefficients are on the same order of magnitude, with
equal precision in the constant terms as for the case $\alpha=2$ where the
exact scaling relation is known. This suggests that the exponents in the scaling form of $1/N$ are
solely dependent on the interaction strength and independent of the order of the \ren entropy.
\begin{table} [H]
\caption{Interaction dependence for $\alpha = 1$. Coefficients for Eq.~\eqref{Eq:fittingform} at different values of $V/J$.}
\centering
\begin{tabular}{rcccc}
\multicolumn{5}{c}{$n=1,\ \alpha = 1$} \\
\toprule
{$V/J$}&{$C_{1+4g}$}&{$C_2$}&{$C_{3+4g}$}&{$C_4$} \\ 
\midrule
$-1.65$& $-$ & $8.55(6)\times 10^{-1}$& $-$& $-10.1(6)$\\
$-1.00$  & $9.011(1)\times 10^{-2}$& $1.45(2)\times 10^{-1}$& $9.3(4)\times 10^{-1}$& $2.4(1)$\\
$1.65$  & $1.144(7)\times 10^{-2}$& $5.93(9)\times 10^{-2}$& $-6.9(2)\times 10^{-1}$& $1.63(6)$\\
$1.75$& $1.317(6)\times 10^{-2}$& $3.0(7)$& $-5.5(1)\times 10^{-1}$& $1.28(5)\times 10^{-2}$\\
\bottomrule
\end{tabular}
\label{tab:03_coeffsalpha1}
\end{table}

\begin{table}[H] 
\caption{Interaction dependence for $\alpha = 1$. Leading order coefficient/constant term in Eq.~\eqref{Eq:fittingform} at different values of $V/J$.}
    \centering
\begin{tabular}{rl} 
\multicolumn{2}{c}{$n=1,\ \alpha = 1$} \\
    \toprule
     {$V/J$}& {$C_0$} \\ \midrule
     $-1.65$ & $0.78592(1)$\\
     $-1.00$ & $0.9021179(9)$  \\
     $1.65$ & $0.8553804(2)$\\
     $1.75$ & $0.8442061(2)$   \\
     \bottomrule
\end{tabular}
    \label{tab:04_constantalpha1}
\end{table}

\begin{figure}[H]
    \centering
    \includegraphics{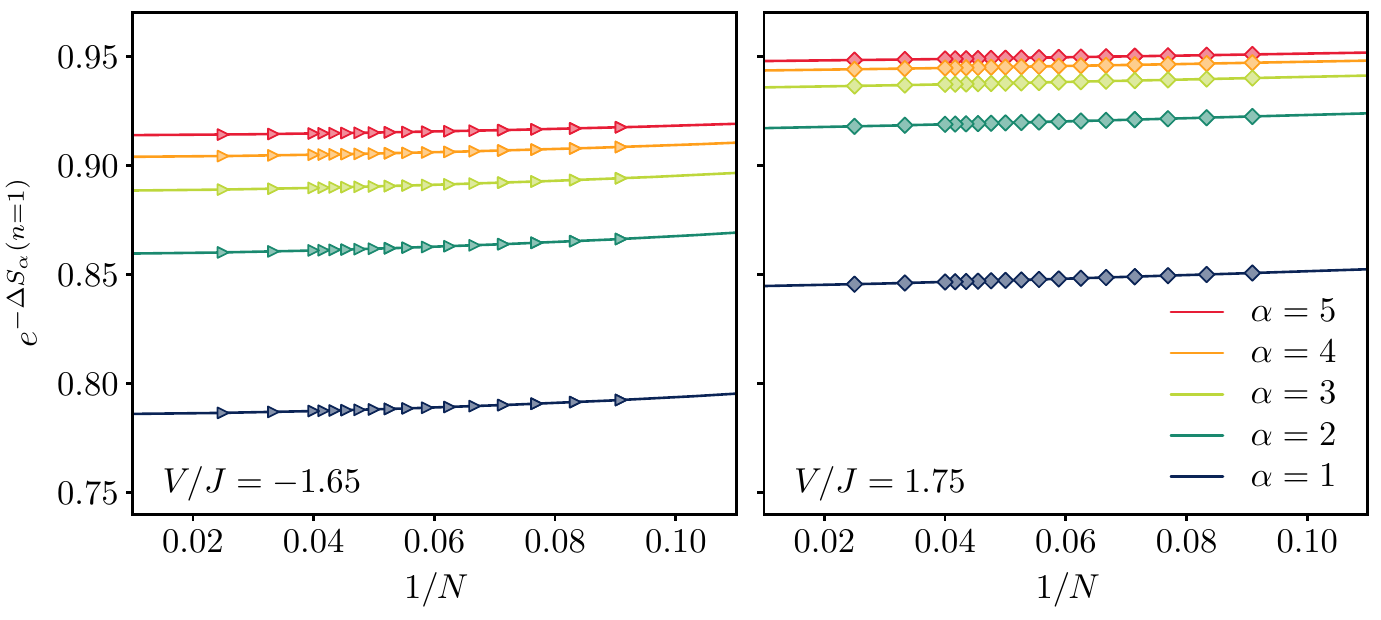}
    \caption{$e^{-\Delta S_\alpha(n=1)}$ for $V/J=-1.65$ (left) and $1.75$
        (right). Solid lines correspond to fitting the data according to Eq.~\eqref{Eq:fittingform}.  The points correspond to data collected from DMRG and ED with different colors used to indicate each $\alpha$. }
    \label{fig:06_v1.75n1}
\end{figure}

Focusing on two fixed interactions $V/J=-1.65,1.75$, the fit to the scaling
form in Eq.~\eqref{Eq:fittingform} is shown for different values of $\alpha$ in
Fig.~\ref{fig:06_v1.75n1}. The coefficients extracted for each $\alpha$ have
the same order of magnitude in their error. The extracted coefficients are
available as data files online \cite{CODEHarini}. This gives numerical
confirmation that the functional form of the scaling of the \ren entanglement
entropy is independent of the \ren index $\alpha$.

\begin{figure}[H]
    \centering
    \includegraphics{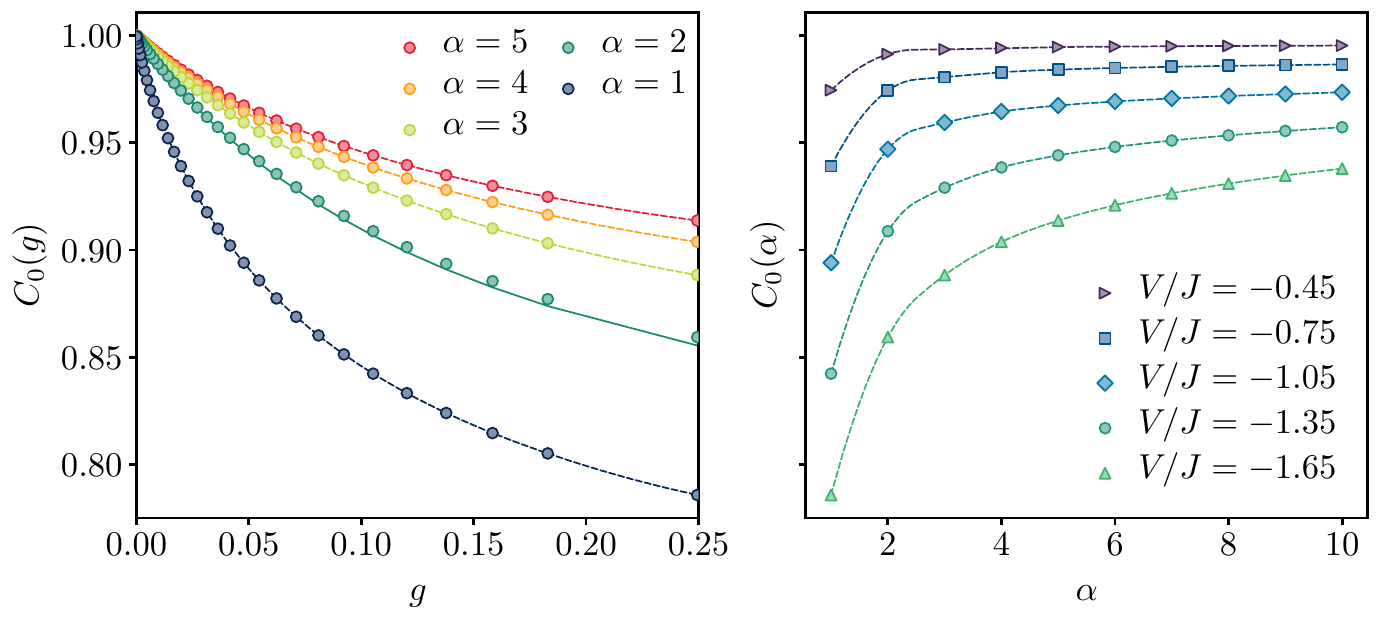}
    \caption{Constant term in fit of $e^{-\Delta S_\alpha(n=1)}$ as a function
    of the interaction $g$ for different $\alpha$. Dotted lines are a guide to
the eye, while the solid line for $\alpha=2$ in the left panel is a prediction from bosonization.}
    \label{fig:07_constants}
\end{figure}

To further understand how the leading order coefficient behaves as a function of interaction strength, we fix $\alpha$ and vary $V/J$. We find that the constant term $C_0$ in Eq.~\eqref{Eq:fittingform_highern} shows similar scaling with the interaction dependent parameter $g$ at each $\alpha$. Similarly, at fixed interaction, we determine the constant term as a function of the order $\alpha$ and see that it exhibits similar scaling for each value of $V/J$. This suggests that unlike the exponents of the $N$ scaling, coefficients depend on both the interaction and the order of the \ren entropy.   The results for both cases are illustrated in Fig.~\ref{fig:07_constants}. 
In the left panel we plot $C_0(g)$ extracted from our non-linear fitting to
Eq.~\eqref{Eq:fittingform} along with a guide to the eye (dashed line) except
in the case $\alpha=2$ where we use a solid line to indicate our definitive
prediction from bosonization given in Eq.~\eqref{Eq:ent_cl_expanded}, again
using $\epsilon$ values as extracted in Ref.~\cite{Thamm:2022}. We observe
excellent agreement for weak interactions with some deviations appearing for
large $g$ as we approach the first order phase transition (we only include $V/J
< 0$ here). This provides further confirmation of Eq.~\eqref{eq:C0galpha2}.

Taken together, the above numerical investigation strongly suggests that the
overall scaling of $\Delta S_\alpha(n=1)$ is consistent with
Eq.~\eqref{eq:Sempirical} even for $\alpha\neq 2$ where bosonization results
are not available, i.e.\@
\begin{equation}
      \Delta S_\alpha(n=1) = -\ln C_0(\alpha,n=1) +\mathcal{O}\left(\frac{1}{N^{\min\{2,1+4g\}}}\right), 
     \label{eq:Sn1alpha}
\end{equation}
where we also observe from the data that the exponent of the leading finite size correction $\gamma_\alpha(n=1)=\min\{2,1+4g\}$ is independent of $\alpha$. 

\subsection{n-Particle entanglement}
\begin{figure}[H]
        \centering
        \includegraphics{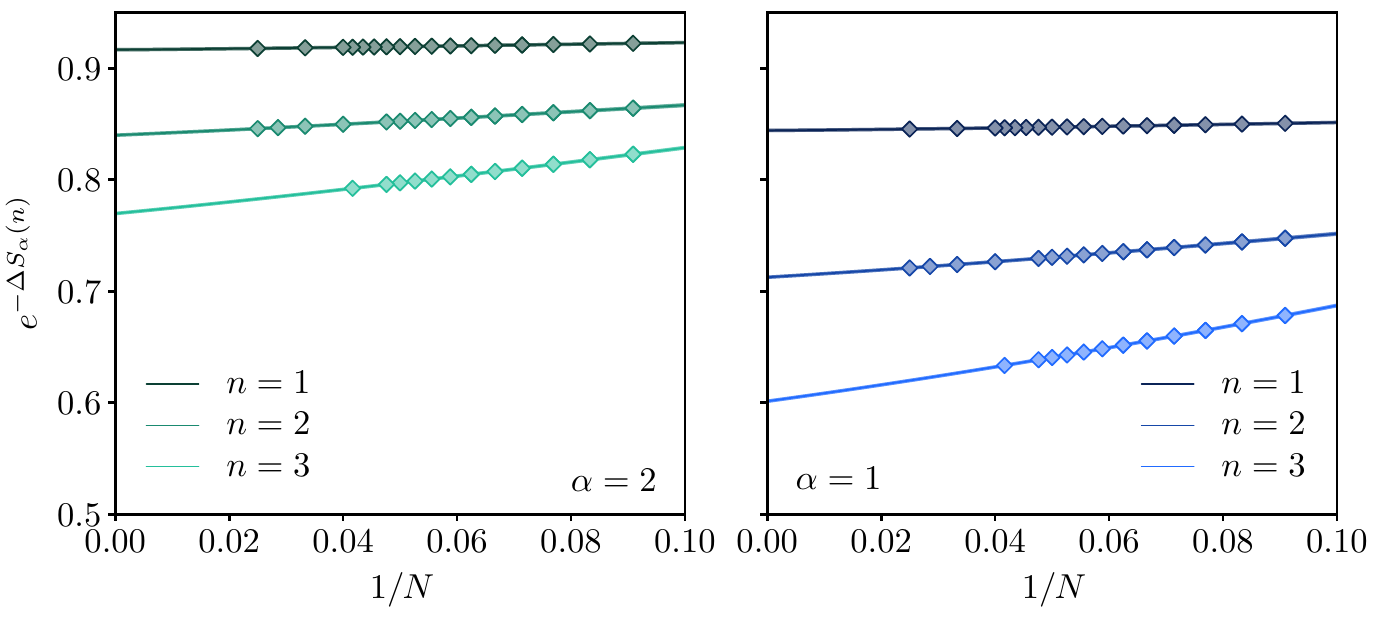}
        \caption{$e^{-\Delta S_\alpha(n)}$ for different bipartitions $n$ and
        $N-n$. Solid lines represent the fit to Eq.~\eqref{Eq:fittingform} for
    $n=1$ and to Eq.~\eqref{Eq:fittingform_highern} for $\alpha=1,2$ and
$V/J=1.75$. Markers represent data collected from ED and DMRG.}
        \label{fig:08_v1.75alpha12}
    \end{figure}

\begin{figure}
    \centering
    \includegraphics{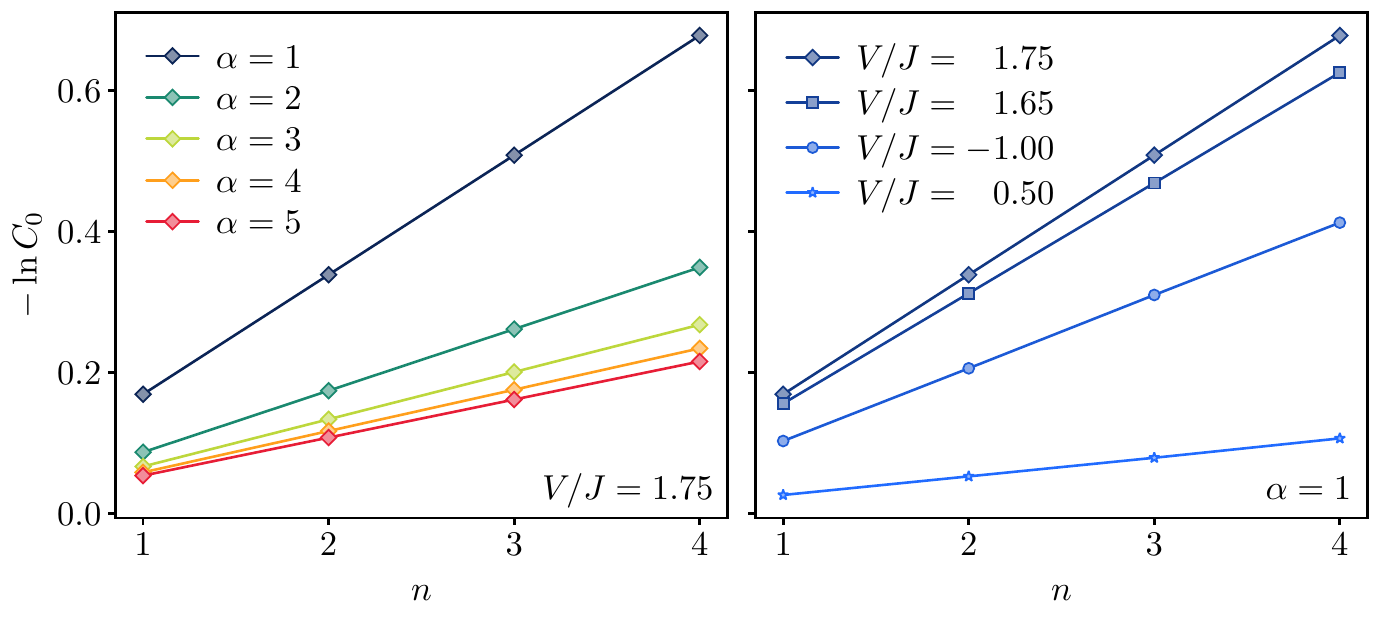}
    \caption{At fixed interaction $V/J=1.75$ and at fixed $\alpha=1$, the subleading constant term of the \ren entropy scales linearly with $n$ for bipartitions of $n=1\ldots 4$ and $N-n$ particles.}
    \label{fig:09_linearityinn}
\end{figure}

Having analyzed numerical results for particle entanglement scaling at a fixed
bipartition of $n=1$ and $N-1$ particles, we now generalize to the behavior of
$e^{-\Delta S_\alpha(n)}$ for $n>1$.  As there are no analytical results to motivate the form of the scaling for $n>1$, we assume a general finite-size expansion holds and focus primarily on the constant term $C_0$:
\begin{equation}
    \begin{aligned}
        e^{-\Delta S_\alpha(n)} &=
        C_0+\frac{C_1}{N}+\frac{C_2}{N^2}+\frac{C_3}{N^3}+\frac{C_4}{N^4}+\dots\, .
    \end{aligned}
    \label{Eq:fittingform_highern}
\end{equation}
Comparisons between this scaling for $n=1,2,3$ are shown in Fig.~\ref{fig:08_v1.75alpha12} for the first and second \ren entropy at $V/J=1.75$ using Eq.~\eqref{Eq:fittingform} for $n=1$ and Eq.~\eqref{Eq:fittingform_highern} for $n=2,3$.
The resulting coefficients for $\alpha=2,1$ are given in Tables \ref{Tab:05alpha2highern} and \ref{Tab:06alpha1highern} respectively.
\begin{table} [H]
\centering
\caption{$n$ dependence for $\alpha = 2$. Coefficients for
Eq.~\eqref{Eq:fittingform_highern} for $n=2,3$ at $V/J=1.75$.}
\label{Tab:05alpha2highern}
\begin{tabular}{cccccc}
    \multicolumn{6}{c}{$\alpha = 2, V/J = 1.75$} \\
\toprule
{$n$} & {$C_0$}& {$C_1$}& {$C_2$} & {$C_3$}& {$C_4$}    \\ 
\midrule

$2$ & $0.839916(5)$&  $0.2163(4)$& $0.906(1)$& $-5.2(1)$& $14.3(6)$   \\
$3$ & $0.769652(6)$& $0.4906(4)$& $1.504(9)$& $-6.55(9)$& $15.5(3)$   \\
\bottomrule
\end{tabular}
\end{table}
\begin{table} [H]
\centering
\caption{$n$ dependence for $\alpha = 1$. Coefficients of Eq.~\eqref{Eq:fittingform_highern} for $n=2,3$ at $V/J=1.75$.}
\label{Tab:06alpha1highern}
\begin{tabular}{cccccc}
\multicolumn{6}{c}{$\alpha = 1, V/J = 1.75$} \\
\toprule
{$n$}&{$C_0$}&{$C_1$} &{$C_2$}&{$C_3$} &{$C_4$}   \\
\midrule

$2$   &$0.712477(5)$&  $0.3071(4)$& $1.25(1)$& $-5.8(1)$&$16.1(6)$ \\
$3$ & $0.601253(6)$& $0.6949(3)$& $2.059(8)$&  $-5.79(2)$& $16.2(3)$   \\
\bottomrule
\end{tabular}
\end{table}
\noindent
We observe that the coefficients can be obtained with the same precision as for $n=1$, which suggests that the scaling form provided by Eq.~\eqref{Eq:fittingform_highern} does not significantly differ from the true form. Particularly, when forcing the exponent $\gamma_\alpha(n>1)$ of the leading correction term to be equal to unity,  we find a high quality of fit over a wide range of interactions, which suggests that the exponent $\gamma_\alpha(n>1)\approx 1$ and has (if any) very weak dependence on the interaction strength. Additionally, we observe that the negative logarithm of the constant term $C_0$ shows a linear dependence on $n$ irrespective of both the interaction strength and order of the \ren entropy as shown in Fig.~\ref{fig:09_linearityinn}. Thus, we predict
\begin{equation}
\ln C_0(\alpha,n)\approx n\ln C_0(\alpha,n=1).
\label{Eq:ScalingC0}
\end{equation}
confirming and characterizing the previously postulated $n$ dependence of the subleading constant, where this result can summarized by the scaling form

\begin{equation*}
    \Delta S_\alpha(n) = -n\ln C_0(\alpha,1)+\mathcal{O}\qty(\frac{1}{N^{\gamma(n)}}) \, 
        \label{Eq:ScalingAtFixed_n},
\end{equation*}
with
\begin{equation}
\begin{split}
\gamma(n=1) &= \min\{2,1+4g\}\\
\gamma(n>1) &\approx 1 \quad .
\end{split}
     \label{Eq:gamma_n}
\end{equation}

\subsection{Construction of a general shape function}

\begin{figure}[h]
\flushleft\hspace{2.5cm}
\includegraphics{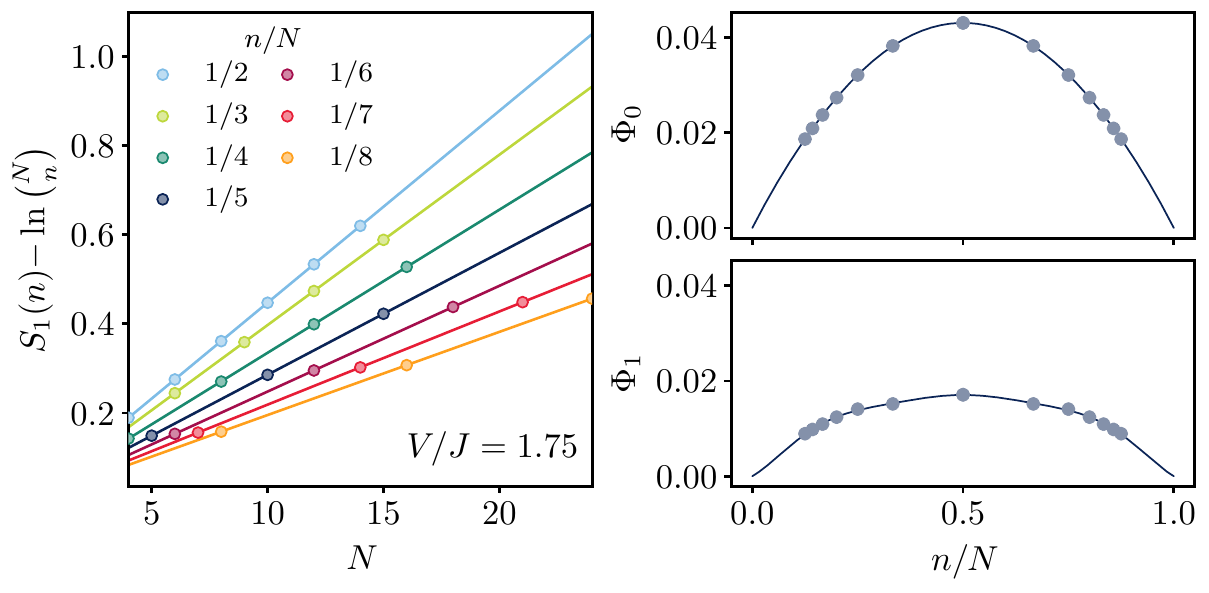}
    \caption{The lefthand panel shows scaling of $S_1-\ln{N\choose n}$ with $N$
    at fixed ratios of $n/N$ at $V/J=1.75$ for $\alpha=1$. Solid lines
correspond to a fit of the data to Eq.~\eqref{Eq:linearscaling} confirming the
linear $N$ scaling.  Markers correspond to data collected from ED and DMRG with
equivalent ratios $n/N$. The righthand panel shows the values of $\Phi_0$ and
$\Phi_1$ for each ratio fitted to the truncated shape function given by Eq.~\eqref{Eq:ffittingformsine}.}
    \label{fig:10_fixedratios}
\end{figure}

Having numerically analyzed the entanglement for $n\le 3$,  we now turn towards
motivating a function for the general $n$-particle scaling that reproduces the
observed asymptotic scaling with $1/N$.  We expect the general form to depend
on both $N$ and $n$, or more conveniently, one can use $N$ and the ratio $n/N$.
Here, we additionally expect $S_\alpha(n)$ to vanish for $n/N =0$ or $1$, and
to be symmetric around $n/N=1/2$, which can be embedded in the expression of
the shape function by assuming it to be a function of $\sin(n\pi/N)$ instead of
$n/N$. Therefore, without loss of generality, we introduce the shape function
$\Phi$ defined via
\begin{equation*}
    \begin{aligned}
        \Delta S_\alpha=S_\alpha(n) - \ln{N\choose n}=N \Phi\qty(\sin \frac{n\pi}{N},N;g,\alpha) \ . 
    \end{aligned}
\end{equation*}
We begin our analysis for the case where $\nu=n/N$ is fixed and $0 < \nu < 1$.  At half-filling the upper bound on $\Delta S_\alpha$ is $\ln{2N\choose n}-\ln{N\choose n}\sim \ln\left[\frac{4\left(1-\nu\right)^{1-\nu}}{\left(2-\nu\right)^{2-\nu}}\right]N$, \emph{e.g.}, if we consider $n/N=1/2$ then the bound scales as $N\ln(\frac{8}{9}\sqrt{3})\approx 0.4315 N$. Therefore, the leading term in $\Delta S_\alpha$ is expected at best to scale linearly with $N$. Thus we can approximate
\begin{equation}
    \begin{aligned}
        \Delta S_\alpha\sim N \Phi_0\qty(\sin \frac{n\pi}{N};g,\alpha) \ , 
    \end{aligned}
\end{equation}
where we have introduced a new simplified shape function with the explicit $N$
dependence removed.  To numerically investigate the scaling of this function,
we separate the data into fixed ratios $n/N = 1/2\ldots 1/8$ and plot $\Delta
S_\alpha$ as a function of $N$ in Fig.~\ref{fig:10_fixedratios}.  
We find that for each ratio, the data collapses onto a straight line suggesting that this function produces a constant at a given $n/N$, an \emph{a posteriori} justification for our assumption of the form of $\Phi_0$. The plot for each ratio is shown in the left panel of Fig.~\ref{fig:10_fixedratios}
where each line is fitted to
\begin{equation}
    \Delta S_\alpha= \Phi_0 N+\Phi_1,
  \label{Eq:linearscaling}  
\end{equation}
where we have introduced a correction term $\Phi_1$, quantifying the deviation
of $\Phi$ from $\Phi_0$.  The resulting coefficients and their corresponding errors are given in Table \ref{tab:fixedratios}.
\begin{table}[H]
    \centering
     \caption{$n/N$ dependence for $\alpha = 1$. Coefficients in Eq.~\eqref{Eq:linearscaling} for fixed ratios of $n/N$ at $V/J=1.75$.}
    \begin{tabular}{lll}
        \multicolumn{3}{c}{$\alpha = 1, V/J = 1.75$} \\
    \toprule
         {$n/N$}&{$\Phi_0$} & $\Phi_1$  \\
         \midrule
         $1/2$&$0.0430(2)$ & $0.01706(3)$\\
         $1/3$ & $0.03821(2)$ & $0.01517(2)$\\
         $1/4$ & $0.03209(4)$ & $0.0141(1)$\\
         $1/5$& $0.02734(3)$ & $0.01094(2)$\\
         $1/6$ & $0.02372(1)$ & $0.0109(2)$\\
         $1/7$&$0.020897(8)$ & $0.0098(2)$\\
         $1/8$&$0.018654(4)$&$0.00893(7)$\\
         \bottomrule
    \end{tabular}
    \label{tab:fixedratios}
\end{table}

With the extensive scaling of $\Delta S_\alpha$ confirmed, we next examine the simplified shape function $\Phi_0$. 
To gain insight into the dependence of $\Phi_0$ on the ratio $n/N$, we expand as
\begin{equation}
\Phi_0\qty(\sin \frac{n\pi}{N};g,\alpha) =\sum_{m=1}^{m_{\rm max}} A_{m}\sin^m\left(\frac{\pi n}{N}\right) \ ,
    \label{Eq:ffittingformsine}
\end{equation}
where we ignore the constant term since $\Phi_0\qty(\sin \frac{n\pi}{N};g,\alpha)$ vanishes for $n/N= 0$ and $1$. By performing a standard non-linear fitting procedure using Eq.~\eqref{Eq:ffittingformsine}, we find that a high-precision fit can be obtained using only the first four terms \emph{i.e.}, $m_{\rm max} =4$
\begin{equation}
\begin{aligned}
    \Phi_0(\alpha,n,N,g) &\approx A_{1}\sin{\left(\frac{\pi n}{N}\right)}+A_{2}\sin^2{\left(\frac{\pi n}{N}\right)}+A_{3}\sin^3{\left(\frac{\pi n}{N}\right)}+A_{4}\sin^4{\left(\frac{\pi n}{N}\right)} \ ,
    \end{aligned}
    \label{Eq:ffittingformsine1}
\end{equation}
as demonstrated by the solid lines in the right-upper panel of Fig.~\ref{fig:10_fixedratios}.  However, to account for the finite size effects we apply the same analysis to the correction term $\Phi_1$ as demonstrated in the right-lower panel of Fig.~\ref{fig:10_fixedratios} and the corresponding fitting coefficients for both of $\Phi_0$ and $\Phi_1$ are listed in  Table \ref{tab:fixedratioshape}.
\begin{table}[H]
    \centering 
    \caption{Values of coefficients in Eq.~\eqref{Eq:ffittingformsine1} for both functions $\Phi_0$ and $\Phi_1$ at $V/J=1.75$ and $\alpha=1$.}
    \begin{tabular}{lllll}
    \multicolumn{5}{c}{$\alpha = 1, V/J = 1.75$} \\
    \toprule
         & $A_{1}$ & $A_{2}$ & $A_{3}$ & $A_{4}$  \\
    \midrule
         $\Phi_0$& $0.0551(2)$ & $-0.021(1)$ & $0.014(2)$ & $-0.0052(8)$\\
         $\Phi_1$ & $0.016(4)$ & $0.04(2)$ & $-0.08(3)$ & $0.04(1)$\\
         \bottomrule
    \end{tabular}
   
    \label{tab:fixedratioshape}
\end{table}

\begin{figure}[h]
        \centering
\includegraphics{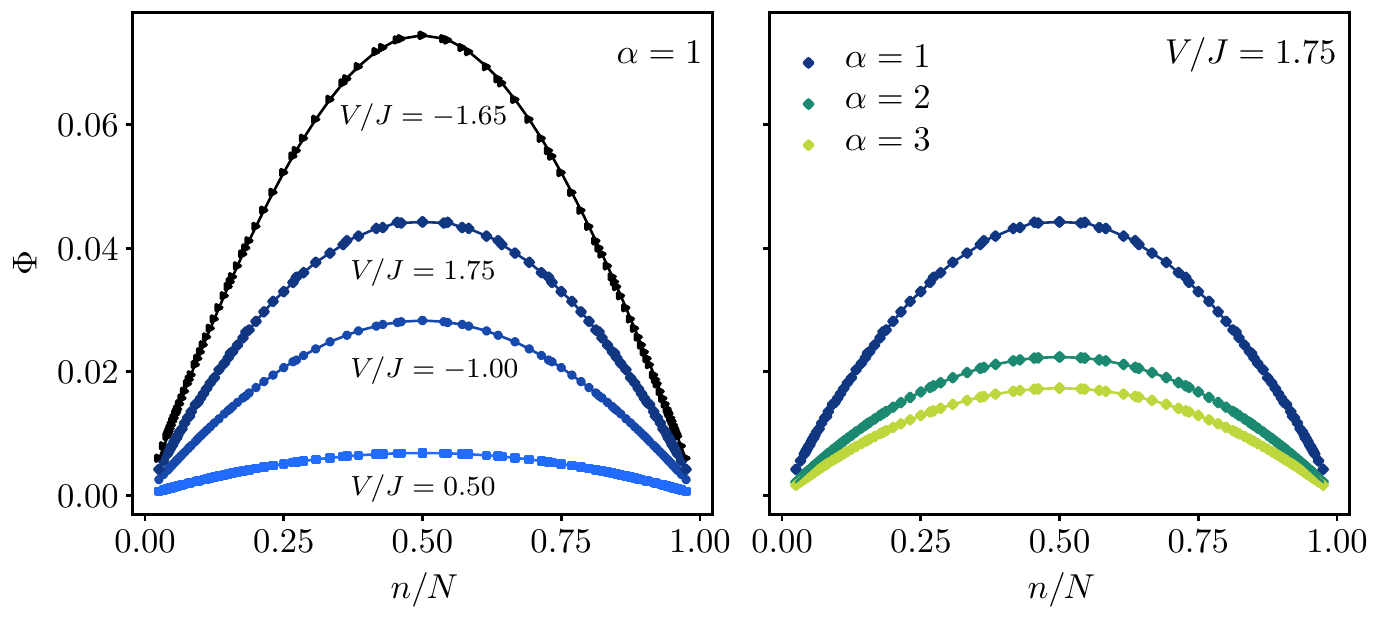}
    \caption{The left panel shows the dependence of $[S_1-\ln{N\choose n}]/{N}$ on $n/N$ at different values of $V/J$ for fixed $\alpha=1$. The right side shows the same quantity at $V/J=1.75$ for different values of $\alpha$. Markers represent data collected from DMRG and ED while solid lines represent a fit to Eq.~\eqref{Eq:ffittingformsine}. The dark blue diamonds for $V/J = 1.75$ and $\alpha=1$ are included in both panels for comparison.}
\label{fig:12_fulldataset}
\end{figure}

We calculate the shape function for a large set of ratios $n/N$ and for $N$ in the range $11\le N\le40$. The results are illustrated in Fig.~\ref{fig:12_fulldataset}, which includes data corresponding to different interaction strengths and \ren indices $\alpha$. For each interaction strength and \ren index in Fig.~\ref{fig:12_fulldataset}, we use the same subset of ratios $n/N$ as in Table \ref{tab:fixedratios} to calculate fitting coefficients for the asymptotic shape function $\Phi_0$ and the corresponding finite-size correction $\Phi_1$. The solid curves represent $\Phi = \Phi_ 0+\Phi_ 1/N$ for each ratio in Fig.~\ref{fig:12_fulldataset}. We find excellent agreement between the fitted curve and the data points, validating the asymptotic shape function $\Phi_0$ for ratios $n/N$  beyond the set in Table \ref{tab:fixedratios}.


\begin{figure}
    \centering
    \includegraphics[width=10cm]{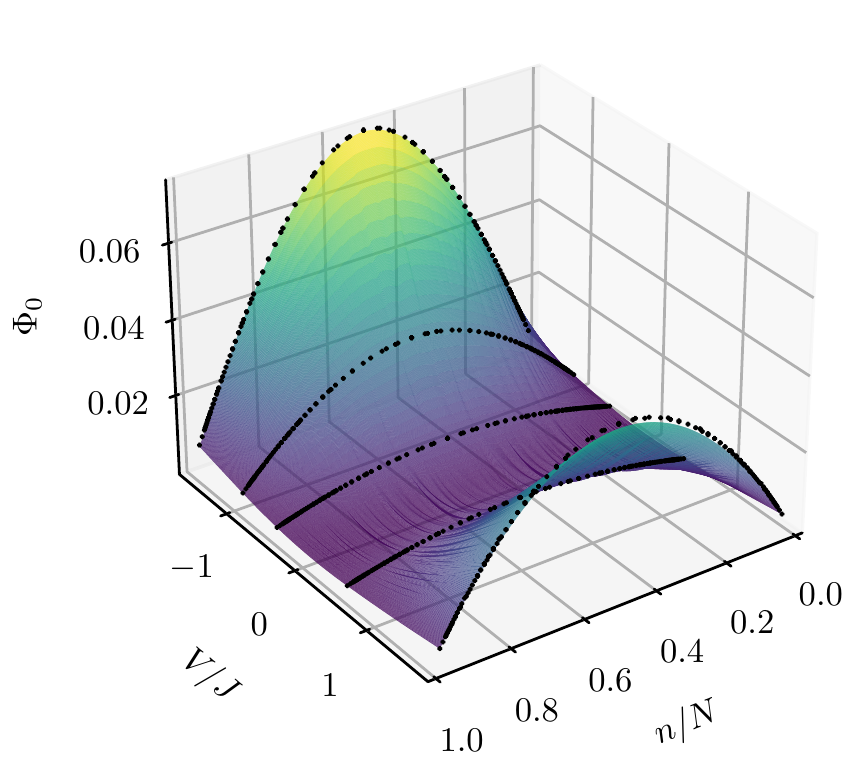}
    \caption{Interpolation of the surface given by $\Phi_0(n/N,V/J)$ along with the data (black circles) for $S_1$ at $V/J=-1.65,-1.00,-0.5,0.5,1.75$. The surface follows the shape of a saddle that reflects the asymmetry of the importance of interactions near the first order and continuous quantum phase transitions.}
    \label{fig:13_interpolatedphi}
\end{figure}

We further compute $\Phi_0(n/N, V/J)$ for each data point at interactions $V/J=-1.65,-2,-0.5,0.5,1.75$ and $\alpha=1$ and interpolate to generate a surface plot of $\Phi_0$. We find that the surface exhibits a saddle shape, with a pommel and cantle that grow with interaction strength. The surface along with the data is shown in Fig.~\ref{fig:13_interpolatedphi}.

To this point, the numerical analysis of the shape function we have carried out suggests that for $N\gg1$, the interaction-induced entanglement of particles follows the general scaling form: 
\begin{equation}
    \Delta S_\alpha\sim N\sum_{m=1} A_{m}\sin^m\left(\frac{\pi n}{N}\right) \ .
     \label{Eq:generalscaling}
\end{equation}
Building on this, we can consider the asymptotic scaling of $\Delta S_\alpha$ for $n\ll N$ by Taylor expanding the sine functions in Eq.~\eqref{Eq:generalscaling}, which yields
\begin{equation}
    \Delta S_\alpha\sim A_{1}\pi n +\mathcal{O}(n^2/N)\ .
     \label{Eq:generalscalingFn}
\end{equation}
This deduced scaling form is consistent with  the result of the previous subsection, \emph{i.e.}, Eq.~\eqref{Eq:ScalingAtFixed_n}, where both relations suggest that as $N\to\infty$, $\Delta S_\alpha$ approaches a constant value that scales linearly with $n$. Comparing the slope of both linear relations suggests that $\ln C_0(n)/n\approx \ln C_0(1)$, which is obtained by fixing $n$ at large $N$, should be equal to $-\pi A_1$, where $A_1$ is obtained from the fitting of the shape function to Eq.~\eqref{Eq:ffittingformsine1}. In order to test this prediction, we extract the value of the coefficient $A_1=-\ln C_0(n)/(n\pi)$ and then compare it with the value of $A_1$ obtained from a direct non-linear fit of the shape function. Table \ref{tab:08_coeffsA} shows this comparison for the case of $V/J=1.75$ and $\alpha=1$, where the value of $A_1$ shows deviations that are on the order of 2\%.
\begin{table} [H]
\caption{Extraction of the $A_{1}$ coefficient for $\alpha=1, V/J=1.75$ through fitting $e^{-\Delta S_1(n)}$ for different values of $n$. Percent differences are determined by comparing the computed coefficients (using $C_0$) to $A_{1} = 0.0551(2)$ extracted from Eq.~\eqref{Eq:ffittingformsine1}.}
\centering
\begin{tabular}{ccccc}
    \multicolumn{5}{c}{$\alpha = 1, V/J = 1.75$} \\
\toprule
{$n$}&{$C_0(n)$}&{$-\ln{C_0(n)}$}&{$A_{1}=-\ln{C_0(n)}/(\pi n)$} & \% Difference \\ 
\midrule
$1$ &  0.8442062(2) & 0.1693585(2) & 0.05390848(8) & 2.2(4)\\
$2$ &  0.712477(6) & 0.339008(8) & 0.053955(1)& 2.1(4)\\
$3$ &  0.601253(6) & 0.508739(9) & 0.053979(1)& 2.0(4)\\
\bottomrule
\end{tabular}
\label{tab:08_coeffsA}
\end{table}

\section{Discussion}
\label{sec:Discussion}

In this paper, we combined bosonization techniques with exact diaogonalization and density matrix renormalization group calculations to study the entanglement entropy between subsets of $n$ and $N-n$ particles in the ground state of the $J$-$V$ model of one dimensional interacting spinless lattice fermions within a delocalized quantum liquid phase. Understanding the scaling of the entanglement $S_\alpha(n)$ as a function of total system size $N$, subsystem size $n$, \ren index $\alpha$ and interaction strength $V/J$ required pushing numerics to unprecedented system sizes, possible only by exploiting all symmetries of the ground state and $n$-body reduced density matrix. The resulting systematic analysis confirms the empirical scaling previously proposed in the literature, and, extends it in two different directions: (i) we identify an interaction induced extensive correction in $N$ at fixed $n/N$ and (ii) determine that in the limit of large $N$, interactions increase entanglement from the non-interacting case by a term that is proportional to the subsystem size $n$. This increase in particle entanglement in an interacting system above its free fermion value (arising for the antisymmetrization of the $N$-particle wavefunction) is encapsulated in a symmetric shape function $\Phi$ which we numerically determine as a function of the ratio $n/N$ and interaction strength $V/J$.  It has a non-trivial saddle-like shape as a result of the fact that both repulsive and attractive interactions between fermions generate non-trivial entanglement combined with the purity of the ground state of the $J$-$V$ model under consideration. 

It is natural to speculate how our results, computed for a 1D critical system, would generalize in the presence of integrability breaking terms in the Hamiltonian, to higher dimensions, or under a change of particle symmetry (e.g.\@ bosons or anyons). For the first, our bosonization calculation would support similar scaling in all critical fermionic systems, and this could be straightforwardly confirmed by adding a next nearest neighbor interaction to the Hamiltonian (see e.g. Ref.~\cite{DelMaestro:2021ja} where the particle entanglement dynamics of a $J$-$V$-$V'$ model was considered).  Eq.~\ref{Eq:motivatedscaling} should also be applicable to itinerant quantum phases of fermions in two and higher dimensions, motivated by the extensive sub-leading behavior of the particle entanglement, unlike the dimension-dependent area law that exists for entanglement between spatial sub-regions. For bosons, the prefactor of the leading order $\ln \binom{N}{n}$ term in the particle entanglement is no longer universal and we leave a systematic study of this case to future work, where it may be possible to study extremely large bosonic lattice models via ground state quantum Monte Carlo \cite{CasianoDiaz:2022sp}.  

The $n$-particle entanglement provides a useful way to quantify quantum correlations generated by interactions encoded in the $n$-particle reduced density matrix, a non-local correlation function that forms a mainstay of many-body methods in both chemistry and physics.  Having detailed knowledge of its finite size scaling behavior and interaction dependence may have practical implications for the measurement and exploitation of this unique form of entanglement as a resource.

\section{Acknowledgments}
This work was supported in part by the NSF under Grant No.~DMR-2041995. 
M.T. and B.R.\@ acknowledge funding by the Deutsche Forschungsgemeinschaft (DFG) under Grant No. 406116891 within the Research Training Group RTG 2522/1 and under grant RO 2247/11-1. 

\appendix

\section{1-Particle entanglement for free fermions}
\label{sec:ffcalc}
We want to directly prove that the second \ren 1-particle entanglement entropy for free fermions is given by $\ln{N}$. Taking the trace of the square of the one body density matrix squared, we have as in Eq.~\eqref{Eq:trace_ff_cl} 

\begin{equation}
    \begin{aligned}
        \Tr[(\rho_1^0)^2] 
        &= \int_{-L/2}^{L/2}dx_2\int_{-L/2}^{L/2}dx_1\frac{1}{N^2}\frac{\sin^2{(\pi N|x_1-x_2|/L)}}{L^2\sin^2{(\pi|x_1-x_2|/L)}} \\
        &= \frac{2}{N^2L}\int_0^{L/2}dx\frac{\sin^2{(N\pi x/L)}}{\sin^2{(\pi x/L)}},
    \end{aligned}
    \label{Eq:theta_ff1}
\end{equation}
where we have used the translational invariance of the system and made the substitution $x=|x_2-x_1|$. Next, we make the substitute $x=\theta L/\pi$, which yields
\begin{equation}
        \Tr[(\rho_1^0)^2]= \frac{2}{N^2\pi}\int_0^{\pi/2}d\theta\frac{\sin^2{(N\theta)}}{\sin^2{(\theta)}}.
    \label{Eq:theta_ff}
\end{equation}
After we simplified the integral we define the function $I(N)=\int_0^{\pi/2}d\theta\frac{\sin^2{(N\theta)}}{\sin^2{(\theta)}}$. We also define another function to be the difference
\begin{equation}
    \begin{aligned}
        J(N) &= I(N+1)-I(N)\\
        &= \int_0^{\pi/2}d\theta\frac{\sin^2\left(\left[N+1\right]\theta\right)-\sin^2{(N\theta)}}{\sin^2{(\theta)}} \\
        &= \int_0^{\pi/2}d\theta\frac{\sin([2N+1]\theta)}{\sin{(\theta)}}.
    \end{aligned}
    \label{Eq:J(N)}
\end{equation}
Again taking the discrete difference of $J(N)$ this time \begin{equation}
    \begin{aligned}
        J(N+1)-J(N) &= \int_0^{\pi/2}d\theta\frac{\sin{([2N+3]\theta)-\sin{([2N+1]\theta)}}}{\sin{(\theta)}} \\
        &= 2\int_0^{\pi/2}d\theta\cos{([2N+2]\theta)} \\
        &=0,
    \end{aligned}
    \label{Eq:J(N)difference}
\end{equation}
and thus
\begin{equation}
 J(N+1) = J(N)=\ldots=J(0).
\end{equation}
We can use this to compute the value of the function $J(N)$ for any $N$, where
\begin{equation}
    \begin{aligned}
        J(0) &= \int_0^{\pi/2}d\theta\frac{\sin{(\theta)}}{\sin{(\theta)}} \\
        &= \frac{\pi}{2}=J(N).
    \end{aligned}
    \label{Eq:J(N)allN}
\end{equation}
Plugging this back into Eq.~\eqref{Eq:J(N)}, we get
\begin{equation}
    I(N+1)-I(N)=\frac{\pi}{2},
\label{Eq:I(N)}
\end{equation}
which immediately dictates that
\begin{equation}
    \begin{aligned}
        I(N) &= I(N-1)+\frac{\pi}{2}=\ldots=I(0)+\frac{N\pi}{2} \\
        I(0) &= \int_0^{\pi/2}d\theta\frac{\sin^2{(N\theta)}}{\sin^2{(\theta)}} = 0 \\
        I(N)&=\frac{N\pi}{2}.
    \end{aligned}
    \label{Eq:I(N)2}
\end{equation}
We can use this in \eqref{Eq:theta_ff}, \begin{equation}
    \begin{aligned}
        \Tr[(\rho_1^0)^2]&=\frac{2}{N^2\pi}I(N) = \frac{1}{N} \\
        \rightarrow S_2(n=1) &= -\ln \Tr[(\rho_1^0)^2] = \ln{N}
    \end{aligned}
\end{equation}
as expected for free fermions. 

\section{Methodology}
\label{app:methodology}
In this Appendix, we provide additional details on the construction of basis states, integer encoding, and computation of the $n$-particle entanglement entropy along with the tricks used to make the computation more feasible for both the exact diagonalization and density matrix renormalization group techniques used to collect the data for our analysis. 
\subsection{Exact diagonalization (ED)}
\subsubsection*{Ground state:}
We first obtain the ground state of Hamiltonian Eq.~\eqref{Eq:jv_Hamiltonian} by 
constructing the ${L\choose N}$ basis states for $N$ fermions on a
lattice of $L$ sites, expressing  the Hamiltonian as a sparse matrix in this basis 
and computing the eigenvector of the smallest eigenvalue using the Lanczos 
algorithm \cite{Lanczos:1950yo}. To efficiently store and operate on the basis states, we 
encode them as an integer fermion basis, i.e., the occupation numbers of a state  are used 
as the bits of the binary representation of an integer: for example  $\ket{1010} \to 10$ and $\ket{1100}\to12$. To further 
reduce the size of the problem, we make use of the symmetries of the Hamiltonian 
Eq.~\eqref{Eq:jv_Hamiltonian}: Translation $T$, which shifts each fermion one site 
to the right on the periodic lattice, e.g.~$T\ket{011001}=\ket{101100}$, is a 
symmetry of the Hamiltonian, $[H,T]=0$, which allows us to group the basis states into
translation cycles. The basis states of a cycle $\nu$ are mapped onto other states 
in the cycle by $T$, where we denote the number of states in cycle $\nu$ as $M_\nu$.
Expressing the Hamiltonian in the basis 
\begin{align}
 \ket{\psi_{\nu, q}} &= \frac{1}{\sqrt{M_\nu}} \sum_{m=1}^{M_\mu} 
 \e^{i\,\frac{2\pi q}{M_\nu}\, (m-1)} T^{m-1}\ket{\varphi_\nu}
\end{align}
allows us to block diagonalize it by sorting the states according to the values of $q$.
Here, $\ket{\varphi_\nu}$ is called the leader of cycle $\nu$ and can in principle 
be any state from the cycle. As the ground state $\ket{\psi_0}$ lies in the $q=0$
block \cite{Barghathi:2022rg}, it is sufficient to only construct and apply the Lanczos algorithm 
to a single block. The $q=0$ translation block can  be further subdivided by a factor of $1/4$
using the inversion and particle-hole symmetry of the $J$-$V$ Hamiltonian \cite{Thamm:2022}. 
\subsubsection*{Particle entanglement entropy:}
For efficiently computing the $n$-particle entanglement entropy, we 
construct the coefficient matrix $C$ of the Schmidt 
decomposition, i.e.\@ we write the ground state $\ket{\Psi_0}$ in terms of the 
basis states $\ket{\theta_a}_A$ and $\ket{\chi_b}_B$ in the $n$ particle partition 
$A$  and the $N-n$ particle partition $B$ as
\begin{align}
    \ket{\Psi} &= \sum_{ab} C_{ab} \ket{\theta_a}_A\otimes \ket{\chi_b}_B\ .
\end{align}
We first construct the sub-bases states in terms of 
translational cycles $\mu$ ($\nu$), i.e.\@ $\ket{\theta_{\mu,i}}$ 
($\ket{\chi_{\nu,j}}$), where $i=1$ corresponds to the cycle leader 
and $T\ket{\theta_{\mu,i}}_A=\ket{\theta_{\mu,i+1}}_A$. Due to the 
translational symmetry, it is sufficient to only consider 
$\ket{\theta_{\mu,i}}_A$ and the cycle leaders in partition $B$, 
$\ket{\chi_{\nu,1}}_B$, as the other contributions are redundant by 
application of $T$. One then needs to decompose the full basis 
states into the Kronecker product of the sub-bases with the appropriate
signs and  extract the corresponding coefficient from the state $\ket{\Psi_0}$.

For this, we pre-compute the so-called structure matrix 
$\mathcal{A}\in\mathcal{M}_{n_{{\rm c},A}\times n_{{\rm c},B}\times L}$, where 
$n_{{\rm c},A}$ ($n_{{\rm c},B}$) is the number of symmetry cycles in 
partition $A$ ($B$). We define the magnitude $|\mathcal{A}_{\mu\nu i}|$ as the index 
of the symmetry cycle for the state $\ket{\theta_{\mu,i}}_A \otimes 
\ket{\chi_{\nu,1}}_B$ in the full basis $\ket{\gamma_{|\mathcal{A}_{\mu,\nu,i}|,k}}$, 
which is also the index for the corresponding coefficient in the vector
$\ket{\Psi_0}$.  The phase of $\mathcal{A}_{\mu\nu i}$ is defined as the sign of 
the state $\ket{\theta_{\mu,i}}_A \otimes \ket{\chi_{\nu,1}}_B$ in the 
anti-symmetrized first quantized basis. Using integer fermion bases, we can
easily construct the integer corresponding to the state in the full basis using
fast bit operations and then look up its position in the ordered list of basis integers.

We choose the convention for the occupation number representation of 
the state that operators act in the reverse order of their corresponding site index from 
$c_L^\dagger$ to $c_1^\dagger$ to determine the sign.  
 The sign of the full state
has two contributions: 
(i) The construction of $\ket{\theta_{\mu,i}}_A$ from the  leader of cycle $\mu$ by
application of $T^{i-1}$ may contribute a minus sign due to the 
boundary conditions if $n$ is even and a fermion is moved 
across the boundary by $T^{i-1}$ as the corresponding creation 
operator has to be moved to the first position through the 
$n-1$ other creation operators. (ii) When combining $\ket{\theta_{\mu,i}}_A$ and 
$\ket{\chi_{\nu,1}}_B$ by acting with the combined creation operators from both 
states on the empty lattice, a sign may arise from ordering these creation operators, e.g.
\begin{align}
    \ket{0001}_A \otimes \ket{1000}_B &\rightarrow c_1^\dagger c_4^\dagger \ket{0} =  \ket{1_2001_1}\\
    \ket{1000}_A \otimes \ket{0001}_B &\rightarrow c_4^\dagger c_1^\dagger \ket{0} = -c_1^\dagger c_4^\dagger \ket{0}= - \ket{1_1001_2} \ .
\end{align} 
Here, the subscript labels the particles. The sign is then given by the parity
of the permutation required for ordering the creation operators.

With this information at hand, we can directly construct the coefficient 
matrix $C$ by combining the coefficients from the state $\ket{\Psi_0}$ 
-- according to the magnitude of the entry in the structure matrix 
-- with the corresponding phase -- according to the sign of the entry in 
the structure matrix. We further exploit translational symmetry and use that
the matrix $C$ can be diagonalized using a Fourier transformation to 
efficiently obtain the singular value spectrum $\lambda_i$.
From the singular value  of the coefficient matrix $C$, we 
obtain the $n$-particle  \ren entanglement entropy of power $\alpha$ as \cite{DelMaestro:2021ja}
\begin{align}
		S_{\alpha}(n) &= \frac{1}{1-\alpha}\,\ln\left(\sum_i |\lambda_i|^{2\alpha}\right)
		\ .
\end{align} 

\subsection{Density matrix renormalization group (DMRG)}
\subsubsection*{Ground state:}
To study even larger systems, we use DMRG as implemented in the \texttt{ITensors.jl}
\cite{Itensor} package to obtain the ground state as a matrix product state (MPS).
In order to stabilize convergence of DMRG to the true ground state, we make sure 
to use a suitable initial state based on the interaction $V/J$, and we ensure that 
the state during the DRMG steps is orthogonal to a subspace of higher translational
cycles as described in Ref.~\cite{Thamm:2022}. 
\subsubsection*{Particle entanglement entropy:}
We compute the $n$-particle entanglement entropy from the $n$-RDM which, in terms of
the ground state $\ket{\Psi_0}$, is given by 
\begin{align}
		\rho_n^{(i_1,...,i_n),(j_1,...,j_n)} 
		= \bra{\Psi} c_{i_1}^\dagger \cdots c_{i_n}^\dagger c_{j_1} \cdots c_{j_n} \ket{\Psi}
		/\textstyle{N\choose n} \ .   
\end{align}
In general, this requires computing ${L\choose n}\times{L\choose n}$ matrix 
elements and constructing a large number of operators, which is slow. However, we can 
utilize several tricks to make computing the $n$-RDM feasible:
\begin{enumerate}[(i)]
		\item   We map the operators $c^\dagger_{i_1}\cdots c^\dagger_{i_n}$ and 
						$c_{j_1}\cdots c_{j_n}$ with $i_1<i_2<...<i_n$ and $j_1<j_2<...<j_n$ 
						onto an integer basis, e.g.\@ $c_1^\dagger c_3^\dagger c_4^\dagger 
						\rightarrow 001101 = 13$. The advantages of the integer basis are 
						that we can efficiently construct  and order the elements 
						such that locating an element in the  
						basis scales only logarithmically with the number of constituents. 
						We can still translate back to the $i_1,...,i_n$ indices to 
						construct the operators in \texttt{ITensors.jl} based on the site 
						indices. 
		\item   We use the anti-commutation relations between the fermion operators 
						to obtain the matrix elements for permutations of the $c^\dagger_{i_1}
						\cdots c^\dagger_{i_n}$ and $c_{j_1}\cdots c_{j_n}$. Here, the parity of 
						the permutation determines  the sign of the corresponding matrix element 
						of $\rho_n$ and the magnitude is the same for 
						all of the permutations. It is even sufficient to only construct the 
						\emph{upper triangle} where $i_1<i_2<...<i_n$, $j_1<j_2<...<j_n$ and to 
						normalize the singular values such that $\sum_n|\lambda_n|^2 = 1$.
		\item   We use translational symmetry of $\ket{\Psi_0}$, i.e.\@ 
						\begin{align}
								\rho_n^{(i_1,...,i_n),(j_1,...,j_n)} 
								&= (-1)^{\delta_{i_n,L}+\delta_{j_n,L}}  
								\rho_n^{(i_1+1,...,i_n+1),(j_1+1,...,j_n+1)} 
						\end{align} 
						to further reduce the number of matrix elements that need to be computed. 
						Here, it is advantageous that we can both efficiently apply the 
						translation operator to the fermion basis states and locate  
						in the basis with logarithmic time complexity.
\end{enumerate}

\bibliography{refs}
\end{document}